\documentclass[12pt,preprint]{aastex}
\usepackage{emulateapj5}

\newcommand{\nh}{N$_{\rm H}$}
\newcommand{\ax}{$\alpha_{\rm x}$}

\newcommand {\sax} {{\it BeppoSAX }}

\newcommand {\sqcm} {cm$^{2}$}

\newcommand {\ergs} {erg~s$^{-1}$}

\newcommand {\chisq} {$\chi ^{2}$}
\newcommand {\rchisq} {$\chi_{\nu} ^{2}$}

\newcommand{\dig}{\hphantom{0}}

\newcommand {\aro} {$\alpha_{\rm ro}$ }
\newcommand {\aox} {$\alpha_{\rm ox}$ }
\newcommand {\amin}{$^\prime$~}

\shorttitle{\sax Observations of Synchrotron X--ray Emission from Radio 
Quasars}
\shortauthors{Padovani et al.}
\begin{document}

\title{\sax Observations of Synchrotron X--ray Emission from Radio Quasars}

\author{Paolo Padovani\altaffilmark{1}} 
\affil{Space Telescope Science Institute, 3700 San 
Martin Drive, Baltimore MD 21218, USA}

\author{Luigi Costamante\altaffilmark{2}}
\affil{Universit\`a degli Studi di Milano, Milano, Italy}

\author{Gabriele Ghisellini}
\affil{Osservatorio Astronomico di Brera, Via Bianchi 46, I-23807 Merate, 
Italy}

\author{Paolo Giommi}
\affil{ASI Science Data Center, c/o ESRIN, Via G. Galilei, I-00044 Frascati,
Italy}

\and

\author{Eric Perlman}
\affil{Joint Center for Astrophysics, University of Maryland, 1000 Hilltop Circle, Baltimore, MD 21250, USA}

\altaffiltext{1}{On assignment from the Space Telescope Operations
Division of the European Space Agency (ESA)}
\altaffiltext{2}{Max-Planck Institute f\"ur Kernphysik, Postfach 10 39
80, D-69029 Heidelberg (current address)}

\begin{abstract}
We present new \sax LECS, MECS, and PDS observations of four flat--spectrum
radio quasars (FSRQ) having effective spectral indices $\alpha_{\rm ro}$ and
$\alpha_{\rm ox}$ typical of high-energy peaked BL Lacs. Our sources have
X--ray--to--radio flux ratios on average $\sim 70$ times larger than
``classical'' FSRQ and lie at the extreme end of the FSRQ X--ray--to--radio
flux ratio distribution. The collected data cover the energy range $0.1 - 10$
keV (observer's frame), reaching $\sim 100$ keV for one object. The \sax band
in one of our sources, RGB J1629+4008, is dominated by synchrotron emission
peaking at $\sim 2 \times 10^{16}$ Hz, as also shown by its steep (energy
index $\alpha_{\rm x} \sim 1.5$) spectrum. This makes this object the {\sl
first} known FSRQ whose X--ray emission is not due to inverse Compton
radiation. Two other sources display a flat \sax spectrum ($\alpha_{\rm x}
\sim 0.7$), with weak indications of steepening at low X--ray energies. The
combination of \sax and ROSAT observations, (non-simultaneous) multifrequency
data, and a synchrotron inverse Compton model suggest synchrotron peak
frequencies $\approx 10^{15}$ Hz, although a better coverage of their spectral
energy distributions is needed to provide firmer values. If confirmed, these
values would be typical of ``intermediate'' BL Lacs for which the synchrotron
and inverse Compton components overlap in the \sax band. Our sources, although
firmly in the radio--loud regime, have powers more typical of high--energy
peaked BL Lacs than of FSRQ, and indeed their radio powers put them near the
low--luminosity end of the FSRQ luminosity function. We discuss this in terms
of an anti-correlation between synchrotron peak frequency and total power,
based on physical arguments, and also as possibly due to a selection effect.
\end{abstract} 


\section{Introduction}
Blazars constitute one of the most extreme classes of active galactic nuclei
(AGN), distinguished by their high luminosity, rapid variability, high ($>
3$\%) optical polarization, radio core--dominance, and apparent superluminal
speeds \citep{kol94,urr95}. 
The broad--band emission in these objects, which extends
from the radio to the gamma--ray band, appears to be dominated by non--thermal
processes from the heart of the AGN, often undiluted by the thermal emission present
in other AGN. Therefore, blazars represent the ideal class to study to further
our understanding of non--thermal emission in AGN.

The blazar class includes BL Lacertae objects, characterized by an almost
complete lack of emission lines, and a subclass of radio quasars (which by
definition display broad emission lines) which have been variously called
highly polarized quasars (HPQ), optically violently variable quasars (OVV),
core-dominated quasars (CDQ). One of their observational properties which is
easier to define is the flat--spectrum radio emission ($\alpha_{\rm r} \la
0.5$) and so we will refer to them as flat--spectrum radio quasars (FSRQ).

Given the lack of prominent emission lines in BL Lacs, more than 95\% of all
known such objects have been discovered either in radio or X--ray surveys.
Follow--up work on radio-- and X--ray--selected samples has shown that the two
selection methods yield objects with somewhat different properties. The energy
output of most radio selected BL Lacs peaks in the IR/optical band 
\citep{gio94,pad95,pad96}; such objects are now
referred to as LBL (low-energy peaked BL Lacs). By contrast, the energy output
of most X--ray selected BL Lacs (referred to as HBL: high-energy peaked BL
Lacs) peaks at UV/X--ray energies.

\citet{pad95} and \citet{sam96} have
demonstrated that the difference in broad-band peaks for HBL and LBL
is not simply phenomenological. Rather, it represents a fundamental
difference between the two sub--classes. The location of the broadband
peaks also suggests a different origin for the X--ray emission of the
two classes. Namely, an extension of the synchrotron emission likely
responsible for the lower energy continuum in HBL, which typically
display steep (energy index $\alpha_{\rm x} \sim 1.5$) X--ray spectra, and inverse
Compton emission in LBL, which have harder ($\alpha_{\rm x} \sim 1$)
spectra \citep{per96,urr96,pad96}. {\it BeppoSAX} observations of BL Lacs are confirming this
picture \citep{wol98,padet01,bec02}.

In this respect, one could expect to find a similar range in peak 
frequencies in the FSRQ class
-- for which, until recently, no evidence existed. Indeed, it was suggested by
some authors \citep{sam96}, based upon the similarities of the
optical--X--ray broad-band spectral characteristics of LBL and FSRQ, that no
FSRQ with synchrotron peak emission in the UV/X--ray band should exist. 

Two studies have drastically changed this picture: 1. the multifrequency
catalog of \citet{pad97} identified more than 50 FSRQ
($\sim 17$\% of the FSRQ in their catalog) spilling into the region
of parameter space once exclusively populated by 
HBL; 2. about 30\% of FSRQ found in the deep X--ray radio blazar
survey (DXRBS) \citep{per98,lan01} were found to have
X--ray--to--radio luminosity ratios, $L_{\rm x}/L_{\rm r}$, typical of HBL
($L_{\rm x}/L_{\rm r} \ga 10^{-6}$ or $\alpha_{\rm rx} \la 0.78$), but
broad (FWHM $> 2,000 {\rm ~km ~s^{-1}}$) and luminous ($L > 10^{43} {\rm ~erg
~s^{-1}}$) emission lines typical of FSRQ. 
The discovery of a large population of ``X--ray strong'' FSRQ (labeled HFSRQ by
\citet{per98} to parallel the HBL moniker) represents a fundamental
change in our perception of the broadband emission of FSRQ. 

X--ray observations of these objects play a fundamental role in finding their
place within the blazar class. For example, if the X--ray spectra were found to
be relatively steep, one could infer a dominance of synchrotron emission, as
observed in HBL. Flatter X--ray spectra, with corroborating evidence from the
whole broad-band emission \citep{pad96}, would instead suggest
inverse Compton emission. In the latter case, the simple equations ${\rm
LFSRQ} \equiv {\rm LBL}$ (where by LFSRQ we mean the ``typical'' FSRQ with
low-energy synchrotron peak) and ${\rm HFSRQ} \equiv {\rm HBL}$ would
not be valid, and some more complicated explanation for the existence of this
class should be sought.

Our previous knowledge of the X--ray spectra of this new class of objects is
scanty and mostly based on low signal--to--noise ratio (S/N) ROSAT data
(Padovani et al. 1997; Padovani et al., in preparation) and is therefore also limited to
the relatively narrow $0.1-2.4$ keV band. The \sax satellite \citep{boe97a}, 
with its broad-band X--ray ($0.1-300$ keV) spectral capabilities, is
particularly well suited for a detailed analysis of the individual X--ray
spectra of these sources.

In this paper we present \sax observations of four HFSRQ candidates, selected
as described below. In \S~2 we present our sample, \S~3 discusses
the observations and the data analysis, while \S~4 describes the results
of our spectral fits to the \sax data. \S~5 discusses the ROSAT data,
\S~6 presents the spectral energy distributions and synchrotron-inverse
Compton fits to the data, \S~7 discusses our results, while \S~8
summarizes our conclusions. Throughout this paper spectral indices are written
$S_{\nu} \propto \nu^{-\alpha}$ and the values $H_0 = 50$ km s$^{-1}$
Mpc$^{-1}$ and $q_0 = 0$ have been adopted.

Readers wanting to skip the details of the data reduction and go directly
to our results can read \S~2, the summaries of our \sax and ROSAT results
in \S\S~\ref{SAX_sum} and \ref{ROSAT_sum}, and then go straight to \S~6. 

\section{The Sample}
The sample selection was done in two separate steps, as our objects were
observed in the \sax Cycle 2 and 3. For both cycles we defined as HFSRQ
flat--spectrum quasars with $\alpha_{\rm ro}$, $\alpha_{\rm ox}$, and
$\alpha_{\rm rx}$ values within $2\sigma$ from the mean values of the HBL in
the multifrequency AGN catalog of \citet{pad97}. These are the usual
effective spectral indices defined between the rest-frame frequencies of 5
GHz, 5,000 \AA, and 1 keV. X--ray and optical fluxes have been corrected for
Galactic absorption. The effective spectral indices have been K-corrected
using the appropriate radio and X--ray spectral indices, available for most
sources (otherwise mean values were assumed). Optical fluxes at 5,000 \AA~have
been derived and K-corrected as described in Padovani et al. (in preparation). We have
chosen this definition so that our selection of HFSRQ matches the traditional
``HBL box'' as detailed in, e.g., \citet{pad95}. 

\centerline{\includegraphics[width=9.0cm]{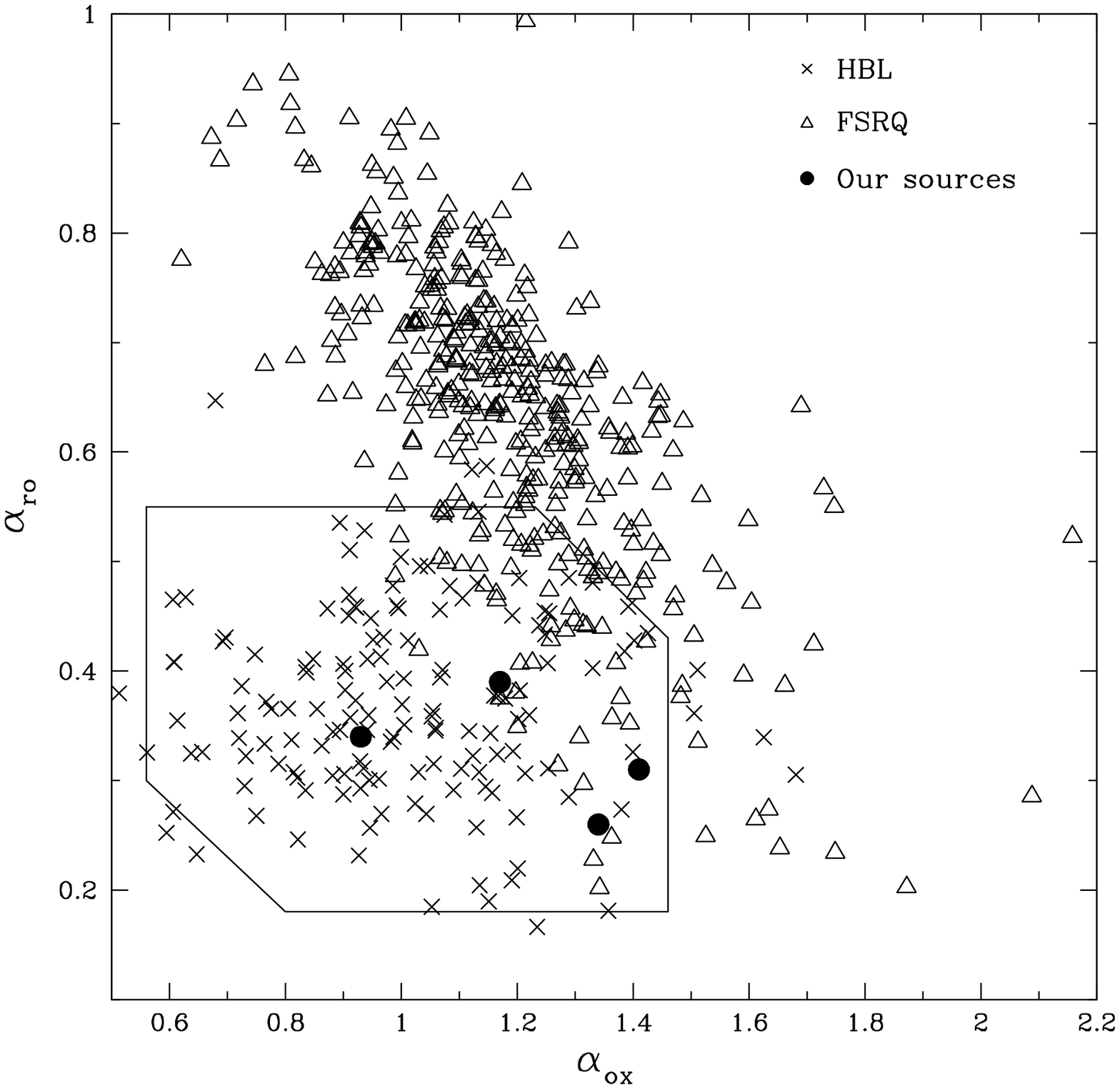}}
\figcaption{The $\alpha_{\rm ro} - \alpha_{\rm ox}$ plane for flat--spectrum
radio quasars (FSRQ) and HBL BL Lacs. Triangles represent FSRQ, crosses
represent HBL, while filled circles represent our sources. The region in the
plane within $2\sigma$ from the mean $\alpha_{\rm ro}$, $\alpha_{\rm ox}$, and
$\alpha_{\rm rx}$ values of HBL is indicated by the solid lines. Data from the
multifrequency AGN catalogue of \citet{pad97}.\label{fig1}}
\centerline{}
\vskip .2in



\begin{table*}
{\footnotesize
\centering
\caption{Sample Properties. \label{tab1}}
\begin{tabular}{lrrlllrrcccl}
\tableline\tableline
Name & RA(J2000) & Dec(J2000) & ~~~$z$ & O & E & F$_{\rm 5GHz}$ & $\alpha_{\rm r}$ & 
$\alpha_{\rm ro}$ & $\alpha_{\rm ox}$ & $\alpha_{\rm rx}$ & Galactic N$_{\rm H}$ \\
 & & & & & & mJy & & & & $10^{20}$ cm$^{-2}$ \\
\tableline
WGA J0546.6$-$6415 & 05 46 41.8& $-$64 15 22&0.323&16.0\tablenotemark{a} &14.7\tablenotemark{b}& 287 &$-0.67$&0.39&1.17&0.66& 4.54 \\
RGB J1629$+$4008   & 16 29 01.3& $+$40 08 00&0.272&17.6 &17.1 &  20 &$-0.29$&0.34&0.93&0.54& 0.85 \\
RGB J1722$+$2436   & 17 22 41.2& $+$24 36 19&0.175&16.5 &15.1 &  35 &$-0.75$&0.26&1.34&0.63& 4.95 \\
S5 2116+81         & 21 14 01.2& $+$82 04 48&0.084&14.6 &13.3 & 376 & $0.26$&0.31&1.41&0.69& 7.41 \\
\tableline

\vspace{-1cm}
\tablenotetext{a}{J magnitude from the US Naval Observatory (USNO) A2.0 catalogue}
\tablenotetext{b}{F magnitude from USNO A2.0 catalogue}

\end{tabular}  }
\end{table*}

\begin{table*}
{\footnotesize
\centering
\caption{\sax Journal of observations. \label{tab2}}
\begin{tabular}{lrcrcrcc}
\tableline\tableline
Name & LECS   & LECS   & MECS & MECS   & PDS & PDS & Observing Date \\
    & exp. (s)   & count rate\tablenotemark{a} (cts/s)   &
  exp. (s)   & count rate\tablenotemark{a} (cts/s)   &
exp. (s)   & count rate\tablenotemark{a} (cts/s)   & \\ 
\tableline
WGA J0546.6$-$6415 & 18564 & $0.036\pm0.002$ & 47234 & $0.053\pm0.001$ & 20004 & $0.148\pm0.059$& 1998 Oct 1-2 \\
RGB J1629$+$4008   & 20512 & $0.033\pm0.002$ & 44759 & $0.023\pm0.001$ & 22798 & $0.086\pm0.054$& 1999 Aug 11-12\\
RGB J1722$+$2436   & 12175 & $0.008\pm0.003$ & 43993 & $0.053\pm0.001$ & 19732 & ...            & 2000 Feb 13-14\\
S5 2116$+$81         & 13171 & $0.147\pm0.004$ & 28826 & $0.218\pm0.003$ & 13296 & $0.286\pm0.073$& 1998 Apr 29\\
S5 2116$+$81      &\dig 5575 & $0.107\pm0.006$ & 19674 & $0.162\pm0.003$ & 10488 & $0.240\pm0.084$& 1998 Oct 12-13\\ 
\tableline
\vspace{-1cm}
\tablenotetext{a}{Net count rate full band}
\end{tabular}
}
\end{table*}

In January 1998, for Cycle 2,  we selected the four X--ray brightest HFSRQ
candidates then known. Three of these came from the AGN catalog of 
\citet{pad97}, while the fourth one was the X--ray brightest HFSRQ in the 
DXRBS
list of \citet{per98}. We were granted \sax observing time for two of
these sources. In January 1999, for Cycle 3, we were able to select even more
extreme candidates by using the ROSAT-Green Bank survey (RGB) 
\citep{bri97,lau98}. Being based on the ROSAT All-Sky
Survey (RASS) the RGB sample has an X--ray flux limit higher than DXRBS
($f_{\rm x} \sim 2 \times 10^{-13}$ erg cm$^{-2}$ s$^{-1}$ vs. $\sim 
{\rm a~few} \times 10^{-14}$ erg cm$^{-2}$ s$^{-1}$) while it reaches
slightly lower 5 GHz radio fluxes ($f_{\rm r} \sim 25$ mJy vs. $\sim 50$
mJy). RGB sources have
then, by selection, higher $f_{\rm x}/f_{\rm r}$ ratios, that is
lower $\alpha_{\rm rx}$, as needed to get more extreme (with synchrotron
peaks, $\nu_{\rm peak}$, at higher energies) HFSRQ. The identification of FSRQ
from the RGB sample is based on a cross-correlation with the NRAO VLA (Very
Large Array) Sky Survey (NVSS) \citep{con98} at 1.4 GHz and is described
in detail in Padovani et al. (in preparation). Amongst all FSRQ in the HBL
region with $\alpha_{\rm rx} < 0.65$ which, based on HBL spectral energy distributions
(SEDs), would correspond to $\nu_{\rm peak} \ga 10^{16}$ Hz \citep{fos98}, we
again selected the four X--ray brightest HFSRQ candidates and, again, were
granted \sax observing time for two out of four sources.
 
The object list and basic characteristics are given in Table 1, which presents
the source name, position, redshift, $O$ (blue) and $E$ (red) magnitudes from
the Automatic Plate Measuring (APM) \citep{irw94}, 5 GHz radio flux, radio spectral
index, $\alpha_{\rm ro}$, $\alpha_{\rm ox}$, and $\alpha_{\rm rx}$ values, and
Galactic \nh.

Figure 1 shows the $\alpha_{\rm ro} - \alpha_{\rm ox}$ plane for FSRQ and HBL
and the location of our sources, well within the region which includes $\sim
95$\% of HBL and clearly offset from where most FSRQ are located. Note
that, in fact, our objects have $\langle \alpha_{\rm rx} \rangle =
0.63\pm0.03$. For comparison, the FSRQ with X--ray data belonging to the 2 Jy
sample ($\sim 80$\% of the total), which include all ``classical''
FSRQ (e.g., 3C 273, 3C 279, 3C 345, 3C 454.3: \citet{pad92}), have $\langle
\alpha_{\rm rx} \rangle = 0.87\pm0.01$, which corresponds to an X--ray--to--radio
flux ratio $\sim 70$ times smaller. We anticipate that our sources have powers 
which, although within the FSRQ range, are relatively low, as discussed in \S\S~
\ref{powers} and \ref{interpr}.

\vskip -2truecm
\centerline{\includegraphics[width=9.0cm]{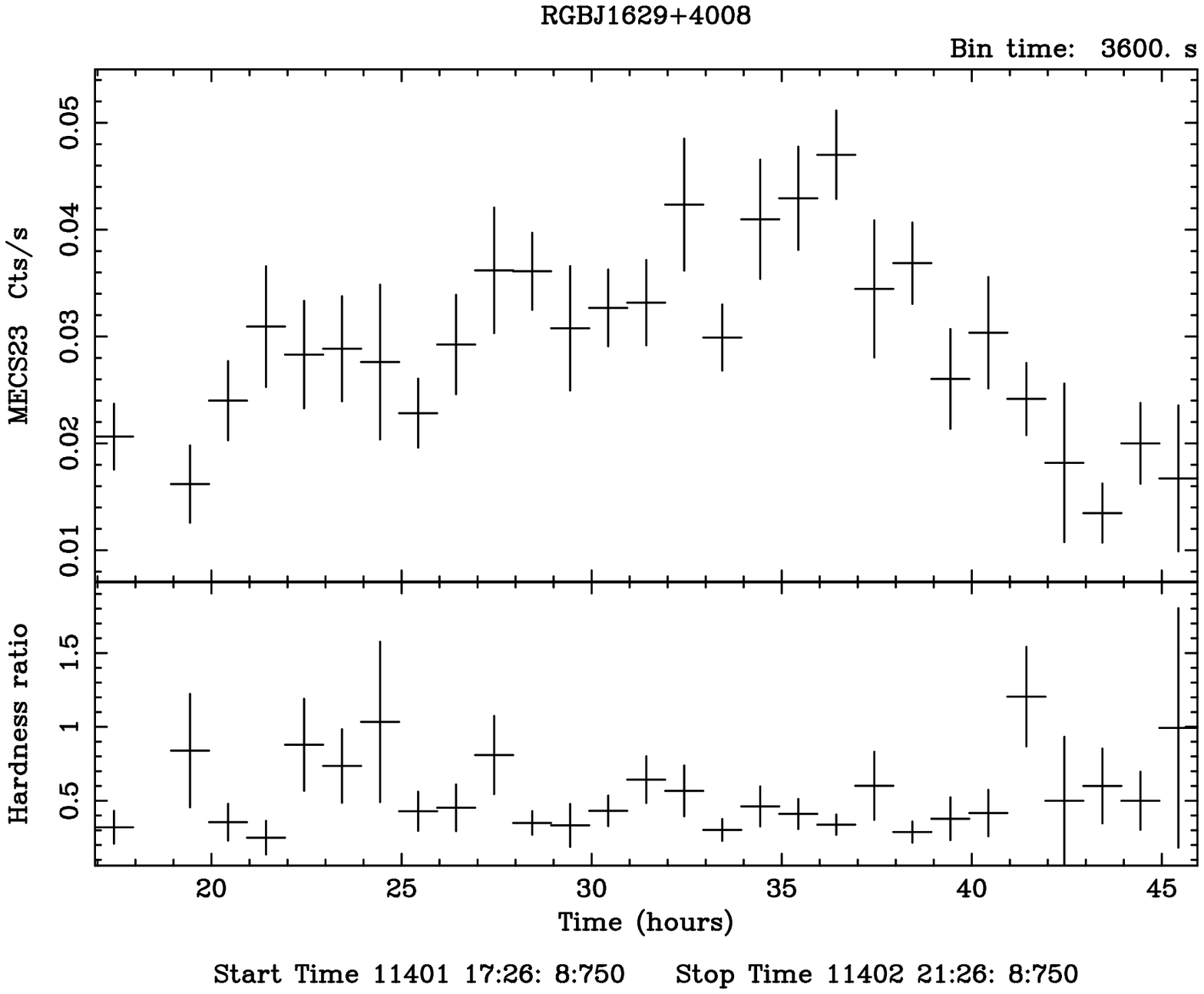}}
\vskip -3.3truecm
\figcaption{The light curve of the two merged MECS units for RGB J1629$+$4008,
along with its hardness ratio, defined as the ratio between the count rates in
the 4--10 keV and 1.5--4 keV bands.\label{fig2}}
\centerline{}
\vskip .2in

\section{Observations and Data Analysis}

\begin{table*}
\vskip -0.1truecm
{\footnotesize 
\centering
\caption{Single power--law fits, LECS + MECS. \label{singpowbl}}
\begin{tabular}{lllllllll}
\tableline\tableline
Name & N$_H$ & $\alpha_{x}$ & $F_{1keV}$ & $F_{[2-10]}$ & $F_{[0.1-2.4]}$ & Norm & 
$\chi^2_{\nu}$/dof  & F--test, notes \\
   & $10^{20}$ cm$^{-2}$ & & $\mu$Jy & erg cm$^{-2}$ s$^{-1}$ & erg cm$^{-2}$ s$^{-1}$ & 
  (L/M) & &  fixed-free N$_{\rm H}$  \\ 
\tableline
WGA J0546.6$-$6415 & 4.54 fixed               & $0.72\pm0.08$ & $0.64\pm0.07$ & 3.84e-12 & 4.14e-12 & 0.82 & 1.07/41  &   \\
                       & $4.60^{+3.59}_{-1.94}$ & $0.72\pm0.10$ & $0.64\pm0.08$ & 3.84e-12 & 4.15e-12 & 0.82 & 1.09/40 & 4\%\\
RGB J1629+4008  & 0.85 fixed              & $1.50\pm0.06$     & $0.66\pm0.05    $ & 1.25e-12 & 7.96e-12 & 0.67 & 0.90/28 &   \\
                     & $0.58^{+0.50}_{-0.43}$ & $1.45\pm0.11$     & $0.63\pm0.06    $ & 1.27e-12 & 7.19e-12 & 0.67 & 0.88/27
		     & 79\%\\
RGB J1722+2436  & 4.95 fixed         & $0.62\pm0.22$          & $0.16\pm0.05 $ & 1.07e-12 & 9.77e-13 & 0.81 & 0.63/18 &   \\
                     & $1.02$ ($<14.6$) & $0.58^{+0.24}_{-0.23}$ & $0.14\pm0.05 $ & 1.08e-12 & 8.85e-13 & 0.79 & 0.55/17 &
		     92\%\\
S5 2116+81         & 7.41 fixed               & $0.73\pm0.04$  & $2.71\pm0.17$   & 1.61e-11 & 1.77e-11 & 0.77 & 0.96/61 & 1998 Apr 29\\
                   & $10.3^{+4.6}_{-3.0}$     & $0.75\pm0.07$  & $2.83\pm0.22$   & 1.60e-11 & 1.87e-11 & 0.78 & 0.91/60 & 96\%\\ 

                   & 7.41 fixed               & $0.77\pm0.07$  & $2.15\pm0.19$   & 1.19e-11 & 1.43e-11 & 0.67 & 0.98/46 & 1998 Oct 12-13  \\
                   & $7.85^{+7.20}_{-3.30}$   & $0.77\pm0.09$  & $2.16^{+0.22}_{-0.19}$ & 1.19e-11 & 1.44e-11 & 0.67 & 1.00/45& 16\%\\ 

        & 7.41 fixed              & $0.74\pm0.04$  & $2.77\pm0.16$     & --- & --- & 0.70 & 1.01/63 & sum  \\
                        & $9.6^{+3.7}_{-2.6}$  & $0.76\pm0.06$  & $2.87\pm0.20$     & --- & --- & 0.70 & 0.97/62 & 92\%\\
\tableline
\vspace{-1.2cm}
\tablecomments{The errors are at $90$\% confidence level for one (with fixed
 N$_{\rm H}$) and two parameters of interest. The fit for S5 2116+81 includes
 PDS data.}
\end{tabular} 
}
\end{table*}

\begin{table*}
\vskip -0.3truecm
{\footnotesize 
\centering
\caption{Broken power--law fits, LECS + MECS. \label{bknbl}}
\begin{tabular}{llllllllll}
\tableline\tableline
Name & N$_H$ & $\alpha_S$ & $\alpha_H$ & E$_{break}$ & $F_{1keV}$ & $F_{[2-10]}$ & 
Norm & $\chi^2_{\nu}$/dof & F--test \\
  & $10^{20}$ cm$^{-2}$ & &  & keV & $\mu$Jy & erg cm$^{-2}$s$^{-1}$ & (L/M) & &  \\ 
\tableline
WGA J0546.6$-$6415  &  $8.70$ ($<20$) &   $1.1^{+1.3}_{-0.3}$ & $0.6^{+0.2}_{-0.5}$ & $2.4^{+3.6}_{-1.0}$  &
		    $0.9^{+0.2}_{-0.1}$ & 3.91e-12 
		    & 0.75 & 0.99/38 & 88\%\tablenotemark{a}, 94\% \\

RGB J1629+4008\tablenotemark{b} & 0.85 fixed   & $1.5(unc.)$ & $1.7(unc.)$ & $3.8(unc.)$ & $0.7(unc.)$  & 1.22e-12  & 0.67 & 0.94/26 & 32\% \\

RGB J1722+2436\tablenotemark{b} & 4.95 fixed   & $1.88$($>0.02$) & $0.6^{+0.2}_{-0.2}$ & $0.9$($<4.7$) & $0.1\pm0.1$  & 1.08e-12  & 0.80 & 0.59/16 & 75\% \\

S5 2116+81         & 7.41 fixed & $0.4^{+0.4}_{-0.6}$ & $0.76^{+0.06}_{-0.06}$ & $1.1^{+2.6}_{-0.3}$  &  $2.7\pm0.3$ & 
1.60e-11 & 0.77 & 0.92/59 & 92\%, 4/29/98 \\
\tableline
\vspace{-1.2cm}
\tablenotetext{a}{Compared to single power--law with free \nh.}
\tablenotetext{b}{Errors are at 90\% confidence level for one parameters of interest.}
\tablecomments{Unless otherwise indicated, errors are at 90\% confidence level 
for two parameters of interest. The values of F--test refer to the comparison 
with single power--laws plus fixed Galactic column densities.}

\end{tabular}
}
\end{table*}

A complete description of the \sax mission is given by \citet{boe97a}. 
The relevant instruments for our observations are the coaligned
Narrow Field Instruments (NFI), which include one Low Energy Concentrator
Spectrometer (LECS) \citep{par97} sensitive in the 0.1 -- 10 keV band;
three (two after May 1997) identical Medium Energy Concentrator Spectrometers
(MECS) \citep{boe97b}, covering the 1.5 -- 10 keV band; and the Phoswich
Detector System (PDS) \citep{fro97}, sensitive in the $15-300$ keV
band, coaligned with the LECS and the MECS. 
A journal of the observations is given in Table 2.

The data analysis was based on the linearized, cleaned event files obtained
from the \sax Science Data Center (SDC) on-line archive \citep{gio97}.
The data from the two MECS instruments were merged in one single
event file by the SDC, based on sky coordinates. The event file was then
screened with a time filter given by SDC to exclude those intervals related to
events without attitude solution (i.e., conversion from detector to sky
coordinates; see \citet{fio99}).
This was done to avoid an artificial decrease in the flux. As recommended by
the SDC, the channels $1-10$ and above 4 keV for the LECS and $0-36$ and
$220-256$ for the MECS were excluded from the spectral analysis, due to
residual calibration uncertainties. 

The spectral analysis was performed using the matrices and blank-sky
background files released in November 1998 by the SDC, with the blank-sky
files extracted in the same coordinate frame as the source file, as described
in \citet{padet01}. Because of the importance of the band below 1 keV
to assess the presence of extra absorption or soft excess (indicative of a
double power--law spectrum), we have also checked the LECS data for differences
in the cosmic background between local and blank-sky field observations,
comparing spectra extracted from the same areas on the detector (namely, two
circular regions outside the $10^{\prime}$ radius central region, located at
the opposite corners with respect to the two on-board radioactive calibration
sources).

We looked for time variability in every observation, binning the data
in intervals from 500 to 4000 s, with null results except for RGB
J1629$+$4008. This source, in fact, clearly varied during the
observation (\chisq-test probability of constancy $<10^{-7}$, see
Fig. \ref{fig2}), doubling its MECS flux in about $\sim15$ hours, with a
more rapid decay of a factor of $\sim 4$ in about 7 hours at the end
of the observation. The same pattern is present in the single MECS
units light curves, and the background showed no significant flux
variations. In spite of the flux variation, no appreciable spectral
changes seem to be present. The hardness ratio, defined as the ratio between
the count rates in the 4--10 keV and 1.5--4 keV bands, is constant
(with a \chisq-test probability $\sim35$\%). We have also looked for
possible differences extracting the spectra in the ``high'' and
``low'' states (defined as $>$ or $<$ 0.03 cts/s, respectively),
finding no significant differences between the spectral indices. 
The LECS data also show evidence of variability, although less significant
than the MECS one (\chisq-test probability of constancy $\sim 5-10\%$,
depending on the bin size). We
stress that our only variable source is also the only one whose X--ray
band is clearly dominated by synchrotron emission (see next section). 

\begin{figure*}[th]
\vskip -1.3truecm
\resizebox{18cm}{!}{\includegraphics[width=9.0cm]{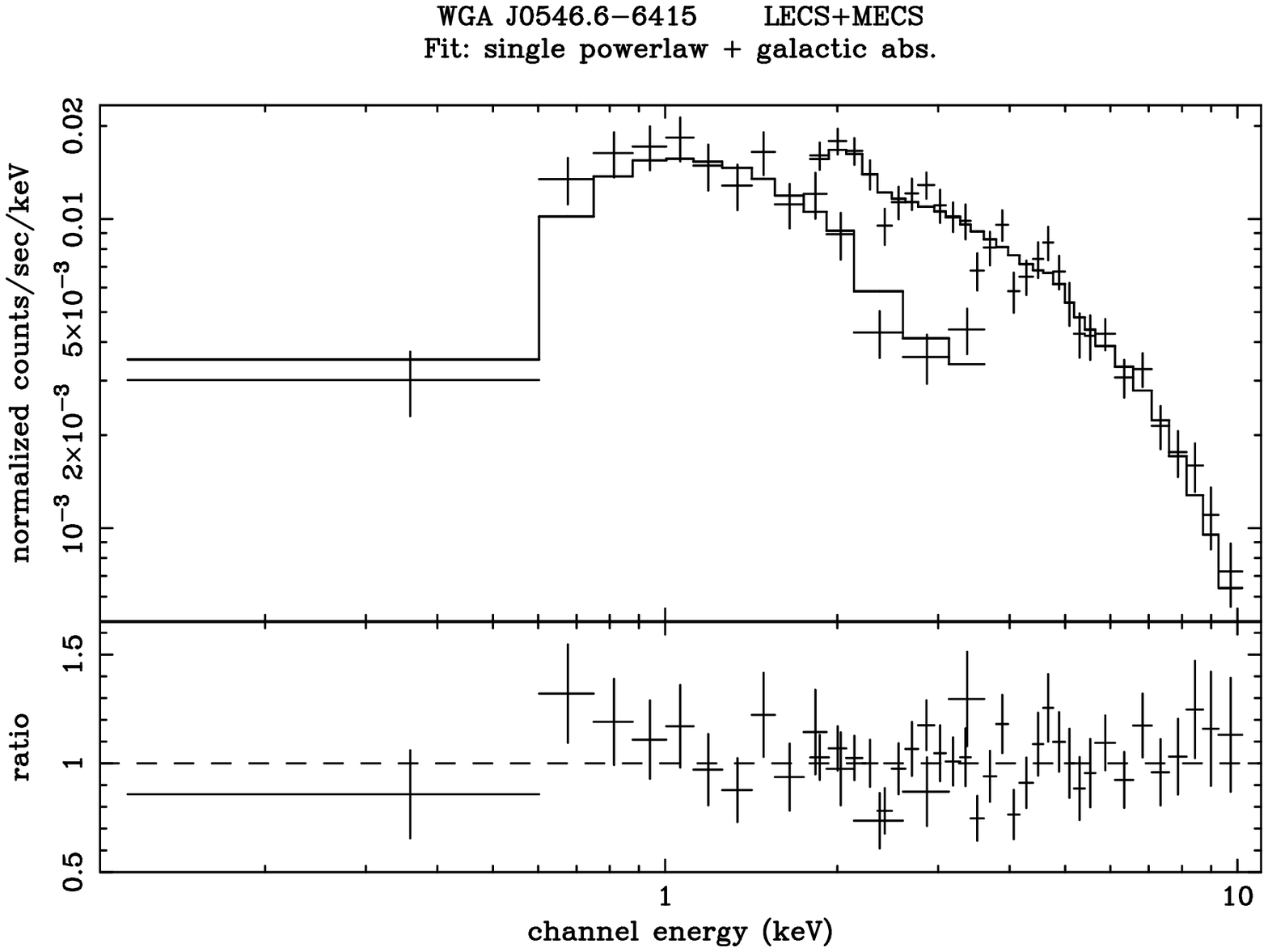}
\hspace*{0.1cm}\includegraphics[width=9.0cm]{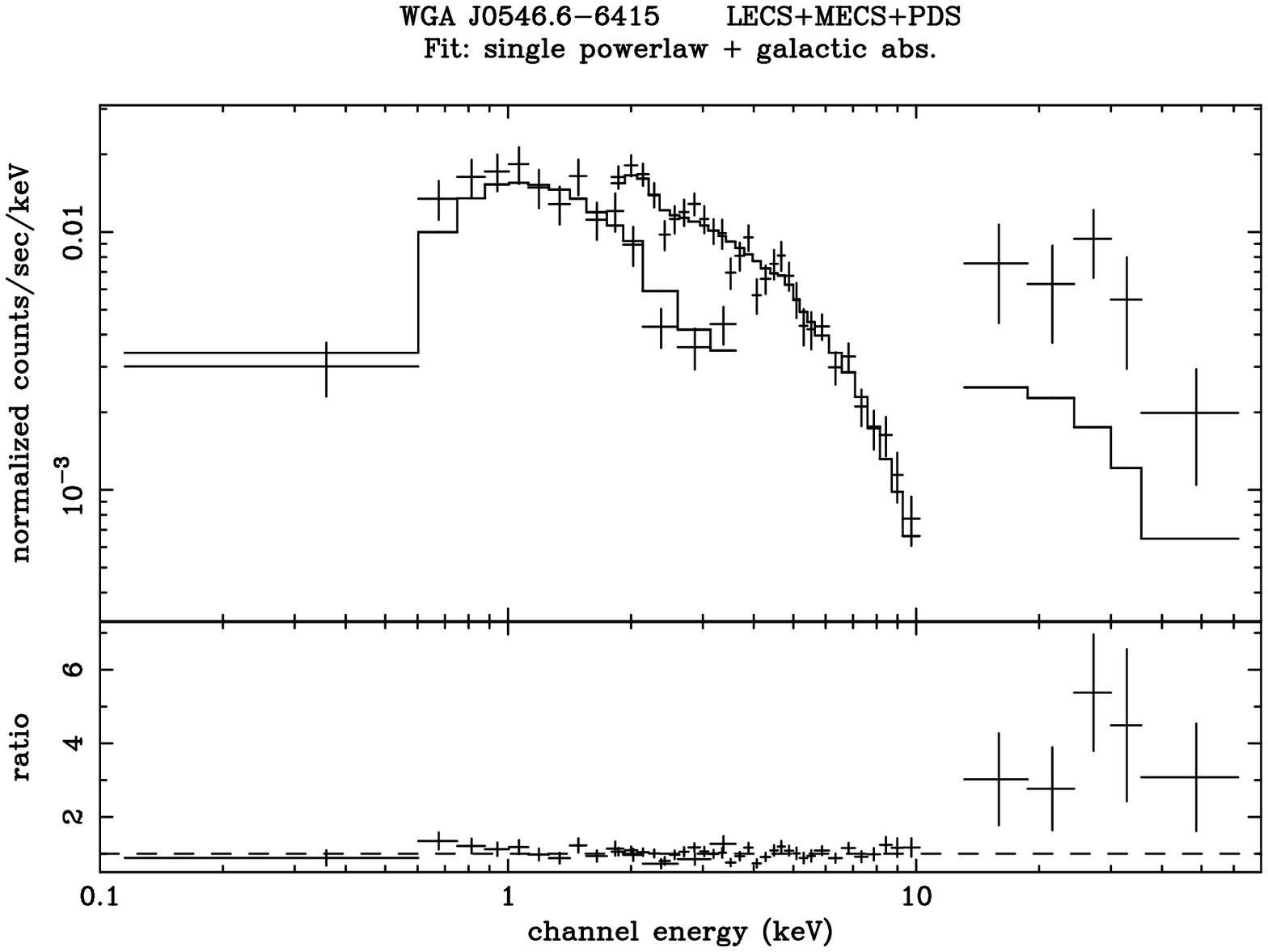}}
\vskip -3.2truecm
\caption{Best fit to the WGA J0546.6$-$6415 data: a single power--law with
Galactic column density.  On the right we also include the PDS data. The PDS
flux is above the LECS+MECS extrapolation due to the likely contribution of
several X--ray sources in the field of view. See text for
details.\label{fig3}}
\end{figure*}

\section{Spectral Analysis}
The spectral analysis was performed with the XSPEC 10.0 package.
Using the program GRPPHA in HEAsoft, the spectra were rebinned with more than 20
counts in every new bin, using the rebinning files provided by the
SDC. Various checks using different rebinning strategies have shown
that our results are independent of the adopted rebinning within the
uncertainties. The LECS/MECS normalization factor was left free
to vary in the range 0.65--1.0, as suggested by the SDC \citep{fio99}.

We fitted the combined LECS and MECS data both with single and broken
power--law models, with Galactic and free absorption. The absorbing column
density was parameterized in terms of N$_{\rm H}$, the HI column density, with
heavier elements fixed at solar abundances and cross sections taken from
\citet{mor83}. The Galactic value was derived from the {\tt nh} program at
HEASARC (based on \citet{dic90}). The \nh~parameter was set free to vary for
all sources to check for internal absorption and/or indications of a ``soft
excess''. The main results of our single power--law fits are reported in
Tab. \ref{singpowbl}. The $F$-test probability quantifies how significant is
the decrease in $\chi^2$ due to the addition of a new parameter (free \nh).
The errors quoted on the fit parameters are the 90\% uncertainties for one and
two interesting parameters, for Galactic and free \nh~respectively. The errors
on the 1 keV flux reflect the statistical errors only and not the model
uncertainties. The results for the broken power--law fit to the data are
presented in Tab. \ref{bknbl}. $\alpha_{\rm S}$ and $\alpha_{\rm H}$ are the
soft and hard energy indices respectively. In this case the $F$-test
probability quantifies the decrease in $\chi^2$ due to the addition of two
parameters (from a single power--law fit to a broken power--law fit). The
errors quoted in this case are the 90\% uncertainties for two interesting
parameters.

We now discuss the details and results of the analysis for each objects.

\subsection{WGA J0546.6$-$6415}
This source is located near the Large Magellanic Cloud (LMC) at
$\sim51\arcmin$ from LMC--X3. Towards the edges of the LECS and MECS images 
(in the region at right ascension $<$ 05$^h$ 46$^m$), apparently
diffuse emission is visible, consistent with the emission of some
sources in the LMC already catalogued through ROSAT PSPC observations
(see \citet{hab99}; hereafter HP99). The most intense source is at
$\sim11\arcmin$ from the quasar, with a count rate $\sim 20$\%
that of the quasar in the MECS images.  It is spatially coincident
with the source [HP99]0049, identified as a foreground star in HP99.
Due to the presence of serendipitous sources we
have used an extraction region of 4\amin also for the LECS instrument.

The data are well fitted by a single power--law model with Galactic
absorption. The best fit plots and values are reported in Fig. \ref{fig3} and
Tab. \ref{singpowbl}.  The obtained spectral index is flat ($\alpha_{\rm x}
<1$), thus indicating an inverse Compton power--law origin of the emission,
given the SED characteristics of this quasar (see \S~6). This result at first
glance does not seem to confirm the ``HBL--like'' nature of this object,
suggested by its location in the \aro$-$\aox plane. We note, however, that a
slightly better fit, even if not significantly improved, can be obtained with
a broken power--law model, allowing for a column density higher than Galactic
(\nh$\sim8.7\times 10^{20}$ \sqcm, $F$-test $\sim88-94$\%, where the two values
refer to a comparison with a single power--law with free and Galactic 
\nh~respectively; see
Tab. \ref{bknbl}). In this case, the steeper component below $\sim2$ keV may
indicate the presence of the tail of the synchrotron emission, as
characteristic of ``intermediate'' BL Lacs (e.g., Giommi et
al. 1999). Variability could also play a role in this (see
\S~\ref{0546_rosat}).

\begin{figure*}
\vskip -2truecm
\resizebox{18cm}{!}{\includegraphics[width=9.0cm]{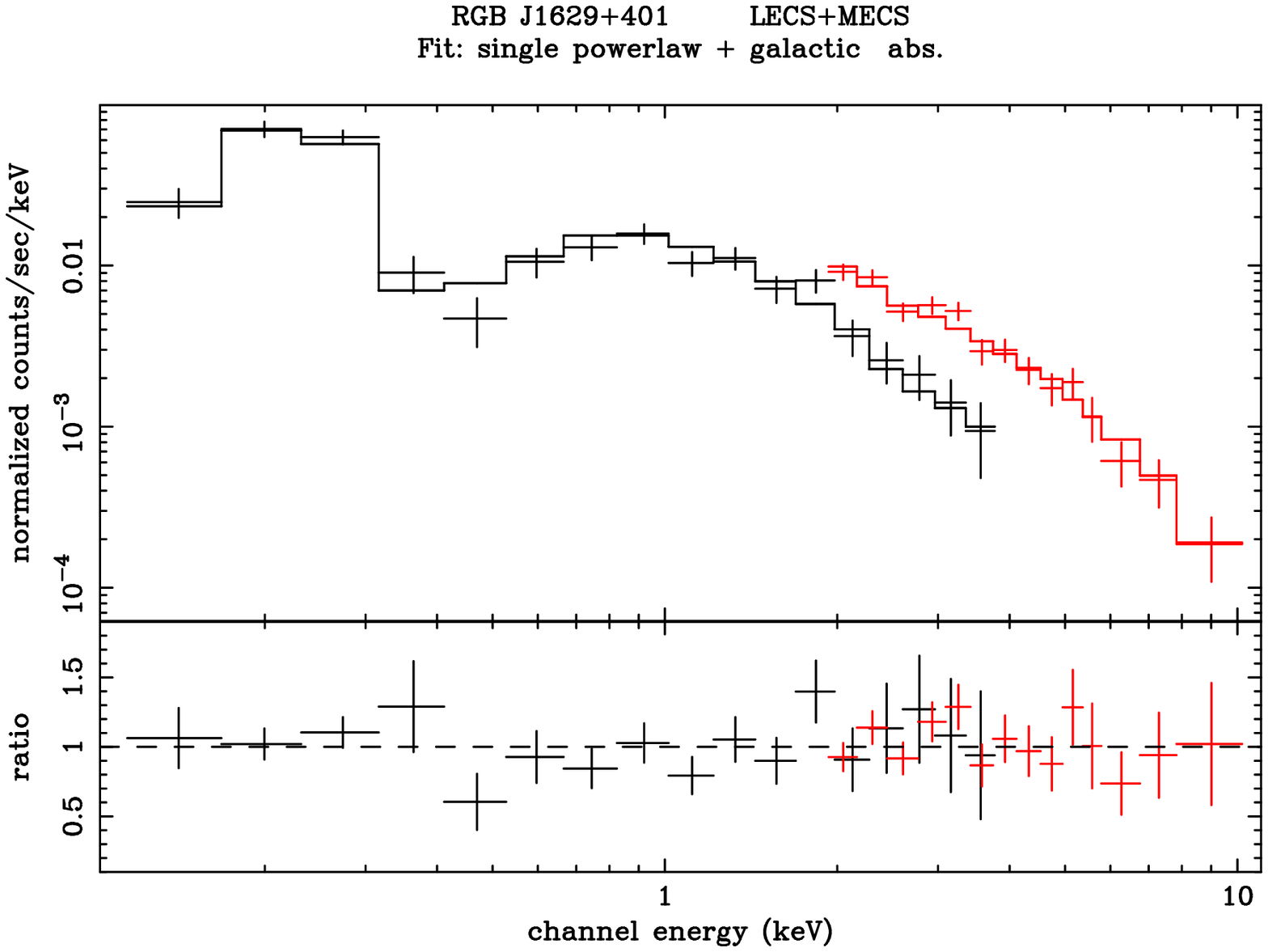}
\hspace*{0.1cm}\includegraphics[width=9.0cm]{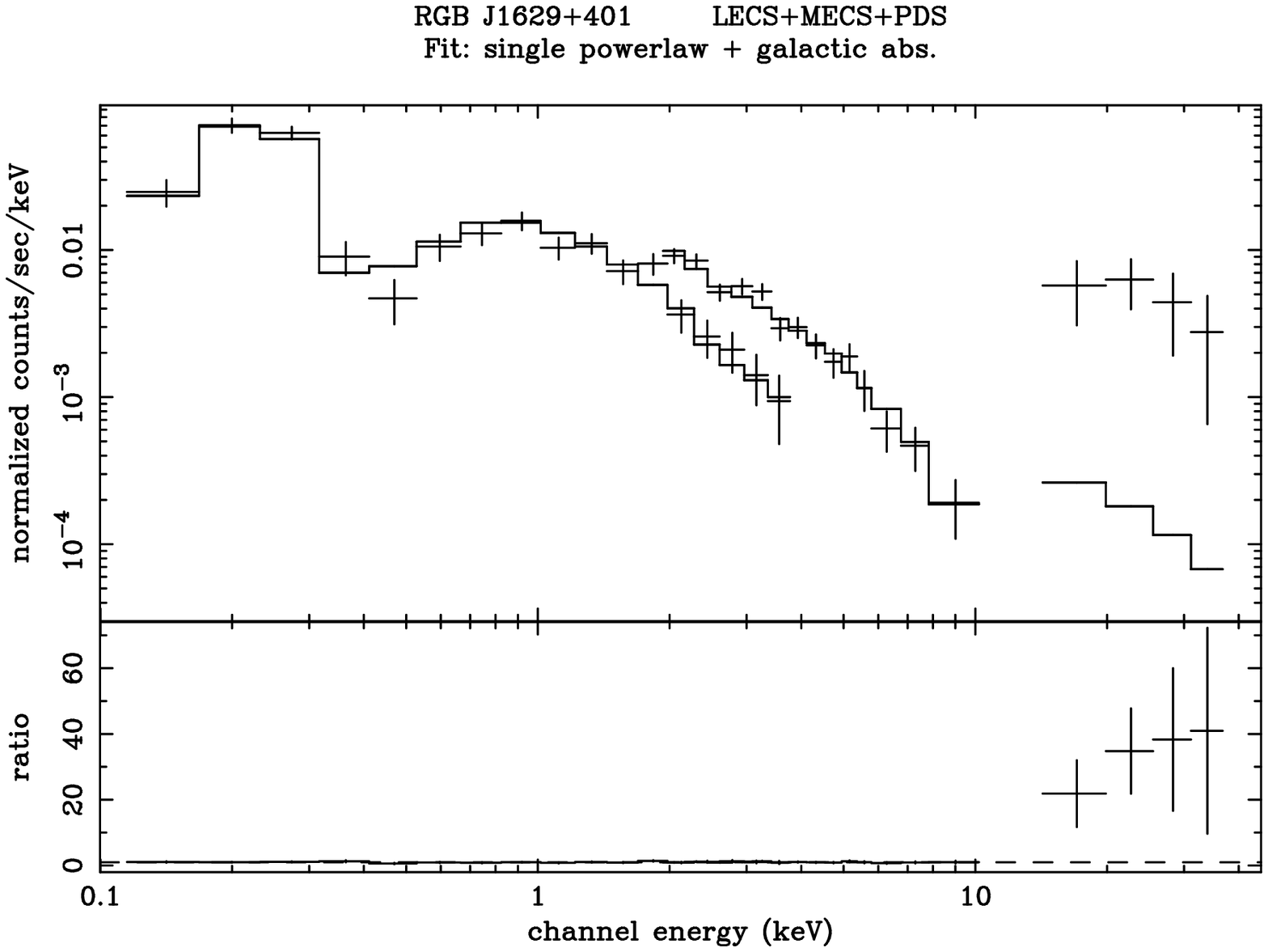}}
\vskip -3.3truecm
\caption{Single power--law fits to the LECS+MECS data of RGB J1629+4008, with
absorption fixed at the Galactic value. On the right we include also the PDS
data. The PDS flux is above the LECS+MECS extrapolation due to the
contribution of A2199. See text for details.\label{fig4}} 
\end{figure*}

Significant flux was detected also in the PDS instrument up to $\sim60$
keV, but with a value not compatible with that expected by a simple
extrapolation of the LECS+MECS spectrum (see Fig. \ref{fig3},
right). Due to the high number of X--ray sources in the field of view,
we consider contamination the most likely explanation for this detection.  
In fact, there
are 9 [HP99] sources within $40\arcmin$ from the quasar, with PKS
0552$-$640, an object classified as AGN in NED, at $40\arcmin$, and
LMC-X3, a well known and bright X--ray binary source, at
$\sim51\arcmin$. Given the known X--ray spectrum of the latter \citep{haa01}, 
we checked its possible contamination level using
the published best fit model, after accounting for the off-axis
efficiency of the instrument. Indeed, the hardest component, a 
power--law of index \ax~$\sim 1.7$ \citep{haa01}, although steep, is quite strong 
($\sim10\mu Jy$ at 1 keV), and provides a non-negligible contribution
up to $\sim35-40$ keV.

\subsection{RGB J1629+4008}
This source is located 37\amin from the center of the well known cluster Abell
(A) 2199. The X--ray emission from the hot gas of the cluster is clearly
visible at the edges of the LECS and MECS images. To avoid contamination
from the cluster, 
therefore, an extraction region of 6\amin has been used for the LECS. The best
fit results are reported in Fig. \ref{fig4} and Tab. \ref{singpowbl}. The
spectrum is well fitted by a single power--law with Galactic values for the
absorbing column density. The resulting spectral index is quite steep
(\ax~$\sim 1.5$), as typical for HBL objects. This source therefore confirms
its ``HBL--like'' properties also in the X--ray spectrum. We stress that RGB
J1629+4008 is the first FSRQ ever found with synchrotron emission extending
all the way to the X--ray band.

The PDS detected this source up to $\sim35$ keV but the flux is
more than an order of magnitude higher than expected from the LECS+MECS extrapolation
(see Fig. \ref{fig4}, right). This is likely due to the cluster emission.
A2199 is a very X--ray bright source (T $\approx 5$ keV, F$_{[2-10]}=1.2 \times 
10^{-10}$ \ergs, emission integrated 
over a region of radius 20\amin; \citet{deg02} and private communication)
and, being at 37\amin from the quasar, is entirely
within the PDS field of view and near where the PDS efficiency is still $\sim50$\% 
(42\amin). 
Although its emission decreases exponentially towards higher energies,
its high brightness still gives a flux in the 15--40 keV band $\sim4.4 \times
10^{-12}$ \ergs,
higher than the quasar LECS+MECS extrapolation ($\sim3.2 \times 10^{-13}$ \ergs).
As a check, we therefore fitted the LECS+MECS+PDS data including in the PDS the best fit model
for the A2199 spectrum \citep{deg02}, after accounting for the off-axis efficiency.
In such a case the PDS data are well accounted for, with a total 
\rchisq$=1.03$ (as compared to 1.38 without the cluster contribution).  

Note that for completeness we also fitted the data for this source with a
broken power--law with absorption fixed at the Galactic value (see
Tab. \ref{bknbl}). No improvement with respect to the single power--law fit
case was obtained.

\vskip -2.3truecm
\centerline{\includegraphics[width=9.0cm]{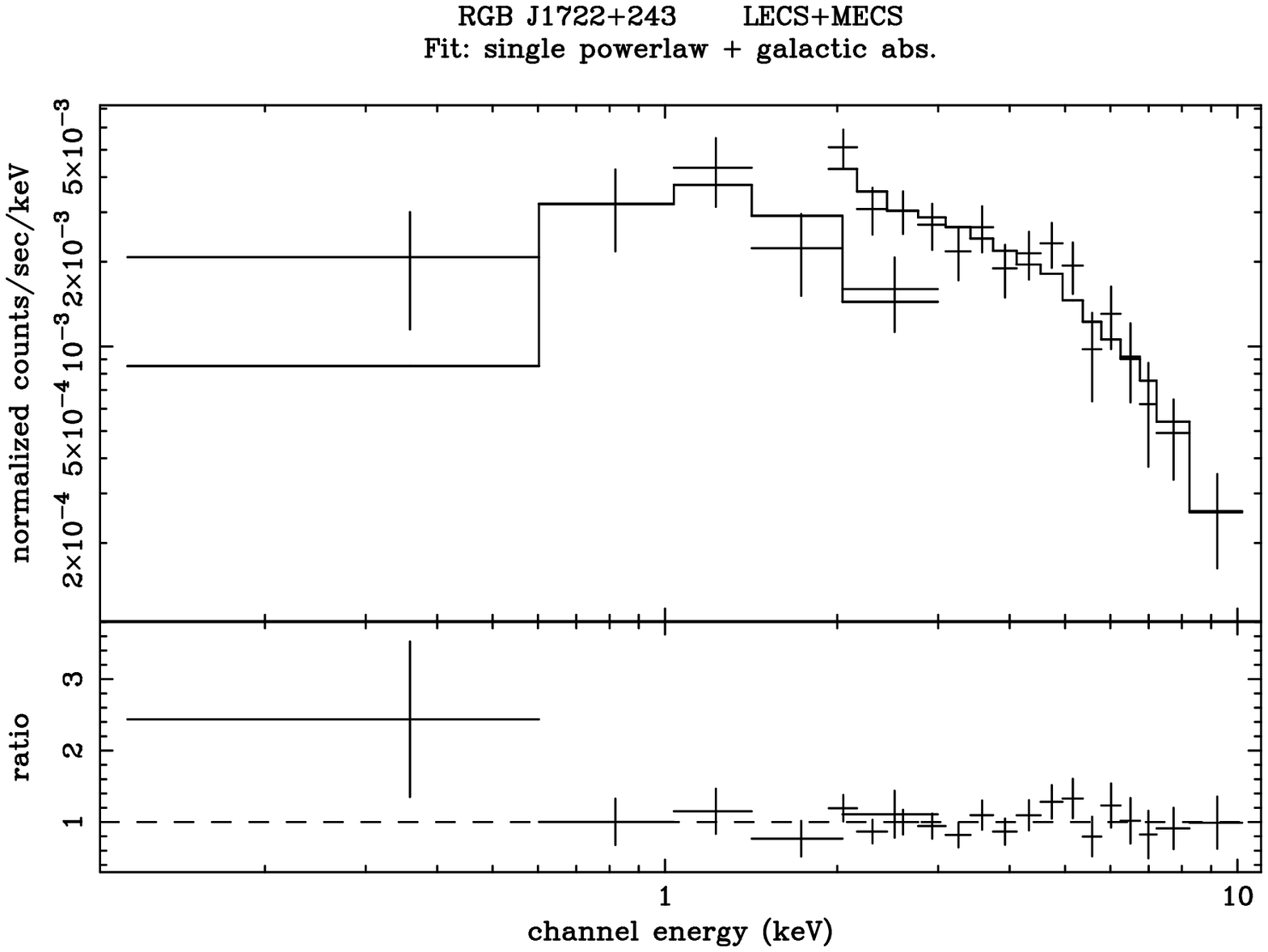}}
\vskip -3.8truecm
\figcaption{RGB J1722+243 single power--law model with Galactic absorption.\label{fig5}} 
\centerline{}
%

\subsection{RGB J1722+2436}

\begin{figure*}[tbh]
\vskip -2truecm
\resizebox{18cm}{!}{\includegraphics[width=9.0cm]{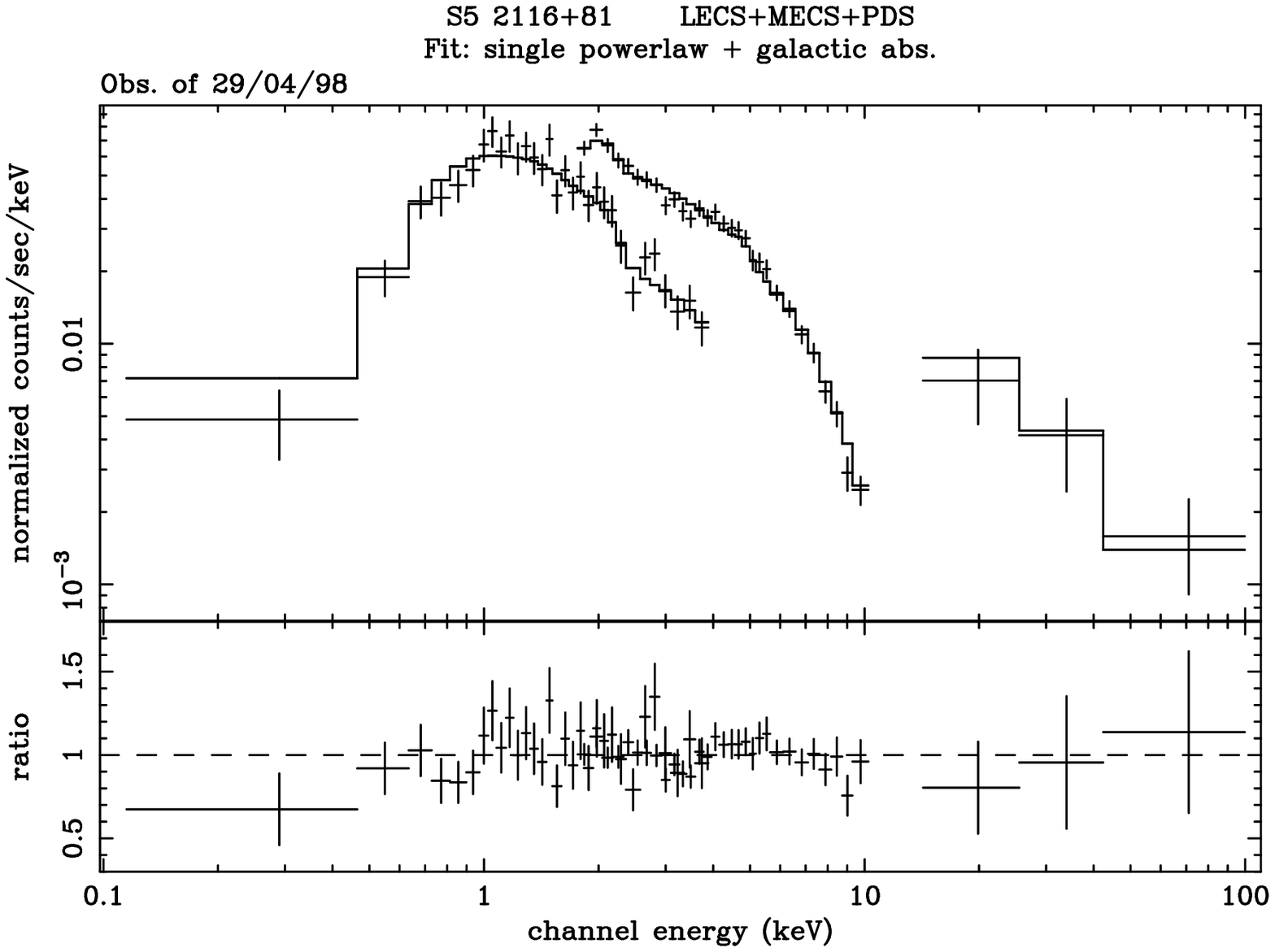}
\hspace*{0.1cm}\includegraphics[width=9.0cm]{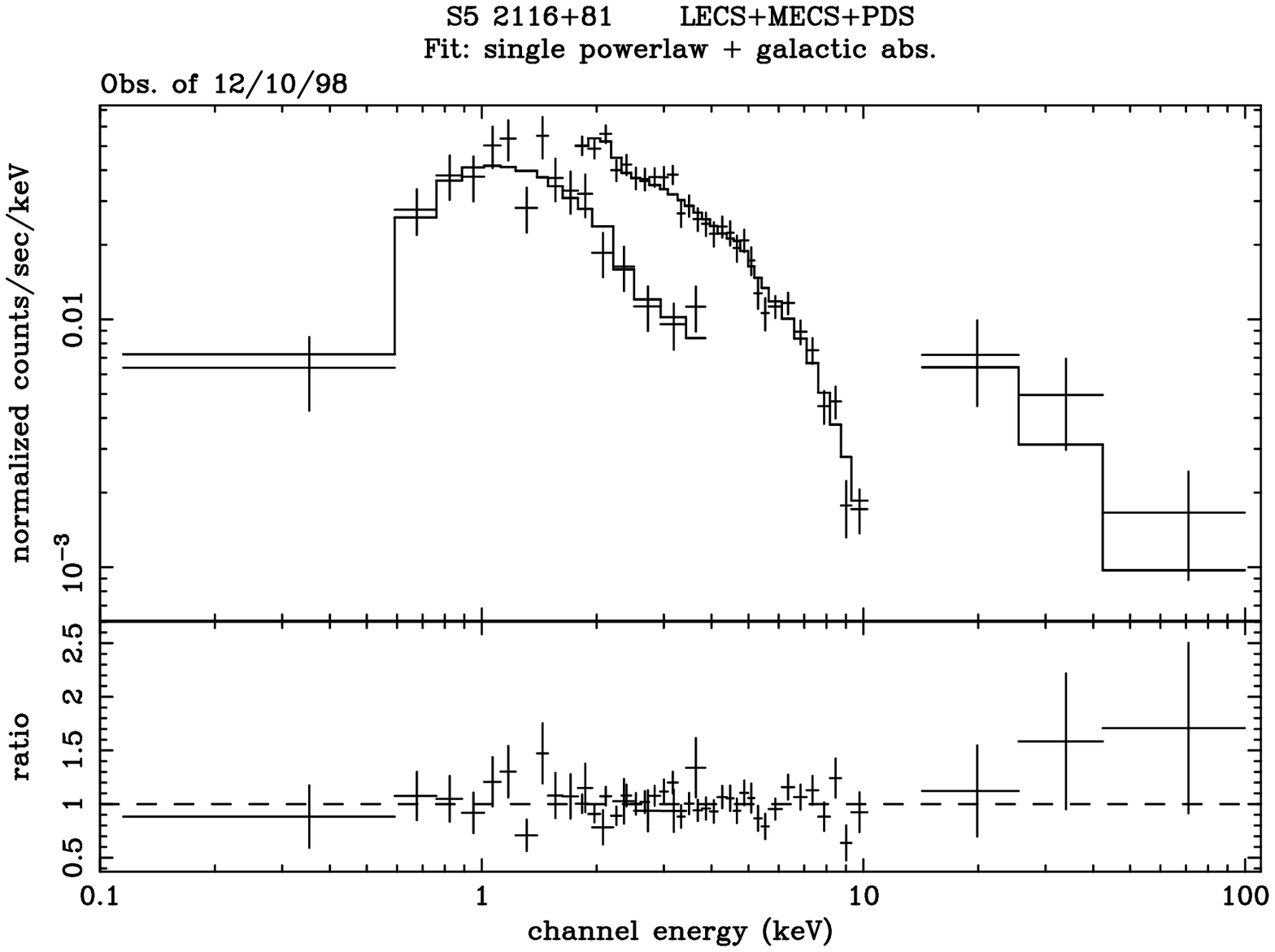}}
\vskip -3.3truecm
\caption{\sax data and fitted spectra for our two observations of S5 2116+81. 
The data are fitted with a single power--law model with Galactic absorption.\label{fig6}} 
\end{figure*}

This source is the weakest of the four observed in the soft X--ray range (see
the LECS count rates in Tab. \ref{tab2}). 
To enhance the S/N, therefore, we
used a 6\amin extraction region for the LECS, as suggested by the SDC \citep{fio99}.
The source is not detected by the PDS.  
The data are well fitted
by a flat ($\alpha_{\rm x} < 1$) single power--law model with Galactic
absorption (see Fig. \ref{fig5} and Tab. \ref{singpowbl}).  The residuals,
however, show the presence of some soft excess below 0.5$-$0.6 keV.  This is
confirmed by the free \nh~fit, which gives values below the Galactic one.  Due
to the low S/N and the narrow band affected, the significance is only marginal
($F$-test $\sim92$\%, last column in Tab. \ref{singpowbl}). A fit with a
broken power--law with the absorption fixed at the Galactic value yields a
steep low-energy component with \ax~$\sim1.9$, though unconstrained (see
Tab. \ref{bknbl}). The improvement with respect to the single power--law fit
and Galactic absorption is however not significant ($\sim75$\%)

The flatness of the X--ray spectrum, as for WGA J0546.6+6415, does not confirm
the ``HBL--like" nature of this objects as suggested by the broad band
indices.  However, the possible steeper component at soft X--ray energies,
if confirmed, may
be attributed to the tail of the synchrotron emission, thus locating the
observed X--ray band in the ``valley" between synchrotron and Compton
emissions (see, in fact, Fig.  \ref{fig7}). In this case, the quasar could be
the FSRQ counterpart of ``intermediate" BL Lacs.  Again, variability could
have also played a significant role.

\subsection{S5 2116+81}
This object was observed by \sax twice, six months apart (see
Tab. \ref{tab2}). In both observations the source was detected in the PDS up
to $\sim90$ keV. The source varied between the two observations
($\sim20$\%), with the highest state in the first observation.  Both datasets
are well fitted by a flat single power--law model (see Fig. \ref{fig6} and
Tab. \ref{singpowbl}), with very similar spectral indices (within the
errors). The PDS data agree, within the errors, with the extrapolation of the
LECS+MECS spectrum.  Given the similarity of the spectral properties, we have
also added the event files to obtain a higher S/N spectrum, whose fit is also
reported in Tab. \ref{singpowbl}.

The free \nh~value agrees with the Galactic one in the second dataset (within
the large errors), while there is a marginal indication of possible excess
absorption in the first and the summed datasets, according to the $F$-test
($96$\% and $92$\%, respectively).  A broken power--law model with Galactic
absorption provides an equivalent fit (giving roughly the same \rchisq~of the
free \nh~fit), with a flatter index (\ax~$\sim 0.4$) below 1.1 keV; however
the significance of the improvement is slightly less ($F$-test $\sim 92$\%;
see Tab. \ref{bknbl}), due to the higher number of parameters.
The flat spectral indices in both observations (i.e., flux states) 
classify this source as a standard FSRQ.

\subsection{Summary}\label{SAX_sum}

To summarize our {\it BeppoSAX} spectral results, the fitted energy
indices are flat, $\alpha_{\rm x} \la 0.8$, for all but one
source. The mean value is $\alpha_{\rm x} = 0.90\pm0.20$ and the
weighted mean is $\alpha_{\rm x} = 1.03\pm0.04$. This latter value is
heavily weighted towards RGB J1629+4008, the only object with a steep
($\alpha_{\rm x} \sim 1.5$) slope. Excluding this source we derive a
mean value $\alpha_{\rm x} = 0.70\pm0.04$ and a weighted mean
$\alpha_{\rm x} = 0.75\pm0.05$.

Three of our sources may have a spectrum which is more complex than a
single power--law, as also illustrated by the residuals to the single power--law
fits. A double power--law model fitted to their spectra indicates, within the rather
large errors, a low-energy ($E \la 1$ keV) excess for WGA J0546.6+6415 and RGB
J1722+2436, with a flatter component emerging at higher energies. In fact, the
spectra are concave with quite a large spectral change, with $\alpha_{\rm S} -
\alpha_{\rm H} \sim 0.5 - 1.3$, and energy breaks around $E \sim 1 - 2$
keV. The hard X--ray spectral index is $\alpha_{\rm H} \sim 0.6$, while
$\alpha_{\rm S} \sim 1.1 - 1.9$. Although the fit is improved by using a
double power--law model, an $F$-test shows that the improvement is more
suggestive than significant, with probabilities $\sim 75 - 94$\%. We do
include RGB J1722+2436 here, despite its relative low $F$-test probability, as
we believe that a best fit $N_{\rm H}$ value a factor $\sim 5$ below the
Galactic one is highly unlikely and strongly suggestive of a soft
excess. We also note that by adding up the $\chi^2$ values 
in Tab. \ref{bknbl}, an $F$-test shows that the improvement in the
fit provided by a double power--law model {\it for the two sources together} is
significant at the $\sim 93$\% level. S5 2116+81, on the other hand,
shows a hint of a {\it flatter} component at low energies ($E \la 1$ keV),
with $\alpha_{\rm S} - \alpha_{\rm H} \sim -0.4$.

\begin{table*}
{\footnotesize
\begin{center}
\caption{WGA J0546.6$-$6415. ROSAT Journal of observations.\label{0546rlog}}
\begin{tabular}{lccrrl}
\tableline\tableline
Obs. ID & Instrument & Off axis & Exposure time &  Net count rate & Observing date  \\
        &            & (arcmin) & (sec.)      &     (cts/s)\tablenotemark{a}&    \\
\tableline
rp140008n00  & PSPC-C & 37.63  &  1334   &   $0.103\pm0.012$  & 1990 Jul 9\\
rp140638n00  & PSPC-C & 37.63  &  1500   &   $0.101\pm0.011$  & 1990 Jul 10\\
rp140009n00  & PSPC-C & 21.98  &  1543   &   $0.124\pm0.011$  & 1990 Jul 9\\
rp140639n00  & PSPC-C & 21.98  &  1628   &   $0.140\pm0.011$  & 1990 Jul 10\\
rp400078n00  & PSPC-B & 51.66  &  7490   &   $0.156\pm0.005$  & 1993 Apr 10\\
\tableline
\vspace{-1.3cm} 
\tablenotetext{}{\hspace{3cm}$^a$ Net count rate full band} 
\end{tabular}
\end{center}  } 
\end{table*}
 
\begin{table*}
{\footnotesize
\begin{center}
\caption{WGA J0546.6$-$6415. ROSAT single power--law fits. \label{0546rfit}}
\begin{tabular}{llllll}
\tableline\tableline
Obs.  & N$_{\rm H}$ & $\alpha_{\rm x}$ & $F_{1keV}$ & $F_{[0.1-2.4]}$ & $\chi^2_{\nu}$/dof  \\
      & $10^{20}$ cm$^{-2}$ & & $\mu$Jy     & erg cm$^{-2}$ s$^{-1}$ &  \\  
\tableline
PSPC-B         & 4.54 fixed  & $1.29\pm0.10$ & $0.88\pm0.06$ & 8.69e-12 & 0.61/31     \\
PSPC-C  sum8   & 4.54 fixed  & $0.8^{+0.2}_{-0.3}$ & $0.45\pm0.05$ & 3.06e-12 & 1.09/16   \\
PSPC-C  sum9   & 4.54 fixed  & $0.8^{+0.2}_{-0.2}$ & $0.48\pm0.04$ & 3.34e-12 & 1.01/14   \\
PSPC-C  fit total  & 4.54 fixed & $0.83^{+0.14}_{-0.15}$ & $0.47\pm0.03$ & 3.22e-12 & 1.00/32  \\   
\tableline
\vspace{-1.2cm}
\tablenotetext{}{\hspace{3cm}Note. --- The errors are at $90$\% confidence level for one  parameter of interest.}
\end{tabular}
\end{center}  }
\end{table*}

In short, one of our sources appear to show an ``HBL--like'' X-ray spectrum,
while the other three are characterized by flat ``LBL--like'' spectra. One, or
possibly two, of these sources, however, show hints of concave X-ray spectra
typical of ``intermediate'' BL Lacs.

\section{ROSAT PSPC Data}

In order to study possible variations of the X-ray spectral properties (and
synchrotron peak) in these objects and
have a better understanding of the underlying emission processes, we have also
looked for other X--ray observations available, in particular from the ROSAT
archive, to take advantage of the higher resolution and collecting area at low
energies. We found Position Sensitive Proportional Counter (PSPC) observations for
two sources, namely WGA J0546.6$-$6415 and RGB J1629+4008. These objects
are located near two famous and well observed X--ray sources, i.e., LMC--X3 and
A2199 respectively, and thanks to the wide field of view of the ROSAT
satellite we could find observations where our targets were present as
serendipitous sources. The journals of the ROSAT observations are reported in
Tab. \ref{0546rlog} and Tab. \ref{1629rlog}. A third object, RGB J1722$+$2436,
was detected in the RASS \citep{vog99}.  We now present the results of the
ROSAT analysis, along with a comparison with the \sax results.

\subsection{WGA J0546.6$-$6415}\label{0546_rosat}

Many PSPC observations of the Large Magellanic Cloud region are present in the
ROSAT public archive, both for B and C detectors, but most have short
exposures ($\la 1000$ s) and/or with the quasar at high off-axis angles.
We therefore decided to analyze the best ones, i.e.,  the longest observation
(7490 s, PSPC-B, pointed on LMC-X3), and those with the lowest off-axis
displacement (PSPC-C). Table \ref{0546rlog} reports the basic information,
namely the observation sequence ID code, the detector used, the off-axis
position of our quasar in the field of view, the exposure time, the net count
rate full band and date of observation. The PSPC-C observations took place in
the same epoch (1990 July 9 and 10), with the satellite often changing
position between the two pointings (thus the two off-axis positions in the
same day). The many short observations have been then reduced and summed
together in a single dataset by the ROSAT data center.
 
We fitted all the PSPC-C datasets independently, with a single power--law
model plus Galactic absorption, obtaining good fits with similar values for
the spectral indices and normalizations, within the errors. To increase the
S/N, therefore, we summed the event files for the two observations with the
same source detector coordinates (i.e., same off-axis angle). The fit results
are reported in Tab. \ref{0546rfit}, where ``sum8'' refers to the summed event
files ..008n00 and ..638n00, and ``sum9'' refers to the summed event files
..009n00 and ..639n00.  Given the similarity of the values, we have also
fitted the two datasets together (sum8 and sum9), using the same model
parameters for both (including normalization).  The resulting fit is good
(\rchisq$=1.00$) thus confirming that both datasets have compatible spectral
properties.


\begin{table*}
\vspace{0.2cm}
{\footnotesize
\begin{center}
\caption{RGB J1629+4008. ROSAT Journal of observations.\label{1629rlog}}
\begin{tabular}{lccrrl}
\tableline\tableline
Obs. ID & Instrument & off axis & exposure time &  net count rate & Observing date  \\
        &            &    (arcmin) & (sec.)      &     (cts/s)\tablenotemark{a} &    \\
\tableline
rp150083n00  &  PSPC-C & 35.28  & 10553  & $0.641\pm0.008$  &  1990 Jul 18\\      
rp800644n00  &  PSPC-B & 35.28  & 40999  & $0.676\pm0.004$  &  1993 Jul 25\\
rp701507n00  &  PSPC-B & 00.29  &  5175  & $0.960\pm0.014$  &  1993 Jul 30\\
\tableline
\vspace{-1.1cm}
\tablenotetext{}{\hspace{3cm}$^a$ Net count rate full band} 
\end{tabular}
\end{center}  }
\end{table*}

\begin{table*}
{\footnotesize
\begin{center}
\caption{RGB J1629+4008. ROSAT single power--law fits.\label{s1629}}
\begin{tabular}{llllll}
\tableline\tableline
Obs.  & N$_{\rm H}$ & $\alpha_{\rm x}$ & $F_{1keV}$ & $F_{[0.1-2.4]}$ & $\chi^2_{\nu}$/dof   \\
      & $10^{20}$ cm$^{-2}$ & & $\mu$Jy     & erg cm$^{-2}$ s$^{-1}$ &  \\  
\tableline
rp150083n00    & 0.85 fixed & $2.03\pm0.05$ & $0.33\pm0.02$ & 8.15e-12 & 2.16/53      \\
               & $0.7\pm0.2$ & $1.95\pm0.14$ & $0.34\pm0.03$ & 7.43e-12 & 2.17/52  \\
rp800644n00    &  0.85 fixed & $2.23^{+0.02}_{-0.04}$ & $0.29\pm0.01$ & 9.68e-12 & 2.12/52   \\
               & $0.82\pm0.01$ & $2.21^{+0.08}_{-0.04}$ & $0.30\pm0.01$ & 9.39e-12 & 2.14/51 \\
rp701507n00   & 0.85 fixed & $2.15^{+0.06}_{-0.05}$ & $0.33\pm0.03$ & 9.48e-12 & 1.37/43  \\
              & $0.4\pm0.2$  & $1.88^{+0.14}_{-0.14}$ & $0.34\pm0.03$ & 6.74e-12 & 0.95/42  \\
\tableline
\vspace{-1.1cm}
\tablenotetext{}{\hspace{2.7cm} Note. --- The errors are at $90$\% confidence level for one (fixed \nh) or two parameters of interest.}
\end{tabular}
\end{center} }
\end{table*}

All datasets analyzed are well fitted by single power--law models with
Galactic column density. The free \nh~fits provide values around the Galactic
ones, and other absorbing models and/or a broken power--law model do not
improve the fits.  The results seem to indicate a variation of the spectral
properties for this source: the 1993 PSPC-B spectrum is characterized by a
steep ``HBL--like'' slope ($\sim1.3$), contrary to what obtained by both the
1990 PSPC-C and 1998 \sax observations, which agree with each other reporting
a flat and roughly equal spectral index ($\sim 0.8$ and $0.7$
respectively). The 1990 and 1998 flux levels are similar, too, with the \sax
flux about 30\% higher than the ROSAT one.  Compared to these, instead,
in 1993 the quasar was in a high state, with a 0.1--2.4 keV flux more than 2.5
times higher than the 1990 values. The X--ray spectrum of this object,
therefore, seems to have a ``steeper when brighter'' behavior, which may
indicate a shift towards higher energies of the synchrotron peak during high
states, when synchrotron emission dominates over inverse Compton in the X--ray
band \citep{pad96,gio99}.

It must be noted, however, that these considerations are based on the 1993
PSPC-B observation, which may be affected by possible calibration problems.
In fact, it has the highest off-axis angle (it is near the edge of the field
of view), and these regions are usually not well studied and calibrated. In
addition, also possible PSPC-B spectral miscalibrations, as reported in
\citet{iwa99} and \citet{min00}, must be taken into
account, as these might steepen the ROSAT slopes by $\sim 0.2 - 0.3$ 
\citep{min00}.

\subsection{RGB J1629+4008}

As for the previous object, we looked in the ROSAT public archive for the best
observations possible, i.e., the longest ones and/or those with the lower
off-axis angle. We found two very long observations of the cluster A2199
taken in 1990 and 1993 with the PSPC-C and B instruments, respectively. These
included our target in the field of view with an offset of $\sim35\arcmin$. We
found also a pointed observation, taken 5 days after the long exposure on
A2199. The basic information for these datasets is reported in
Tab. \ref{1629rlog}.


\begin{table*}
{\footnotesize
\begin{center}
\caption{RGB J1629+4008. Other fits to the ROSAT datasets.\label{b1629}}
\begin{tabular}{lllllllll}
\tableline\tableline
Name & N$_H$ & $\alpha_S$ & $\alpha_H$ & E$_{break}$ & $F_{1keV}$ & $F_{[0.1-2.4]}$ 
& $\chi^2_{\nu}$/dof & F--test \\ 
 & $10^{20}$ cm$^{-2}$ &  &  & keV & 
 $\mu$Jy & erg cm$^{-2}$ s$^{-1}$ & &  \\ 
\tableline
rp150083n00  & 0.85 fixed  & $2.3^{+0.2}_{-0.1}$ & $1.0^{+0.4}_{-0.5}$ & $0.7\pm0.2$ & $0.23\pm0.05$  &  8.70e-12  & 1.47/51 & $>99.9$\%\\
rp800644n00  & 0.85 fixed  & $2.33^{+0.09}_{-0.07}$ & $1.4^{+0.4}_{-0.7}$ & $0.8\pm0.2$ & $0.25\pm0.02$ & 9.95e-12 & 1.43/50 & $>99.9$\%\\              
rp701507n00  & 0.85 fixed  & $2.7^{+3.8}_{-0.4}$ & $1.9^{+0.2}_{-0.3}$ & $0.3^{+0.2}_{-0.1}$ & $0.13\pm0.08$ & 1.08e-11 & 0.91/41 & $>99.9$\% \\   	  
\tableline
     &      &  N$_{\rm H}$ $_{warm}$  & $\alpha_x$  &  E$_{window}$ & $F_{1keV}$ & $F_{[0.1-2.4]}$ 
& $\chi^2_{\nu}$/dof & F-test\\
     &      &   $z=0.03$ &       &    keV       &   $\mu$Jy & erg cm$^{-2}$ s$^{-1}$ &     &    \\   
\tableline
rp150083n00  & 0.85 fixed  &  $11\pm4$ & $1.75\pm0.09$ &  $0.30^{+0.18}_{-0.02}$ & $0.54\pm0.06$ & 8.77e-12 & 1.17/51 & $>99.9$\% \\
rp800644n00  & 0.85 fixed  &  $6.5^{+3.3}_{-2.4}$ & $2.07\pm0.07$ & $0.49^{+0.07}_{-0.27}$ & $0.38\pm0.03$ & 9.78e-12 & 1.27/50 &  $>99.9$\% \\
rp701507n00  & 0.85 fixed  &  $1.6^{+3.5}_{-1.2}$ & $2.2^{+0.2}_{-0.1}$ & $0.25^{+0.03}_{-0.06}$ & $0.36\pm0.05$ & 1.10e-11 & 1.04/41 &  $\ga$99.8\% \\
\tableline
\vspace{-1.1cm}
\tablecomments{Unless otherwise indicated, errors are at 90\% confidence level
for two parameters of interest. The values of the F--test refer to the
comparison with single power--laws plus fixed Galactic column densities.}
\end{tabular} 
\end{center}
 }
\end{table*}

At first we fitted all datasets with a single power--law model, both with
Galactic and free absorption, but these did not provide an adequate fit (see
\rchisq~in Tab. \ref{s1629}). The presence of a more complex spectral shape is
also indicated by the free \nh~ fit of the on-axis observation (rp701507n00),
which provided a good \rchisq~only with absorption values below the Galactic
ones.

We therefore tried alternative models, reported in Tab. \ref{b1629}.  The fits
were significantly improved both adopting a broken power--law model with
Galactic absorption (top three rows in Tab. \ref{b1629}), and adding a warm
absorption component, at the redshift of the A2199 cluster, to the single
power--law model (bottom three rows in Tab. \ref{b1629}). This under the
hypothesis that, at $\sim35\arcmin$ from the center, there is still a not
negligible amount of cluster gas, likely partially ionized.  The model used
for the warm absorber is ZWNDABS in XSPEC, an approximation to the warm
absorber using the photoelectric cross-sections in \citet{bal92}.

The results are a bit puzzling and do not settle clearly the question of which
model should be preferred. The F-test does not help here, given that the
number of model parameters is the same. The off-axis observations seem to be
better fitted by the warm absorber model, according to the \chisq-test
(15-20\% against 2\%, see the \rchisq~in Tab. \ref{b1629}), but this is not
confirmed by the on-axis observation, which has the best \rchisq~for the
broken power--law model. Moreover, the obtained values for the extra (warm)
absorption are different (at the 90\% level), a fact which would imply a
variable absorption with timescales less than five days (the time interval
between the last two observations, i.e., rp800644 and rp701507), quite
unlikely.  For this reason, and possible feature calibration problems for the
off-axis observations (outside the PSPC central rib ring), we consider the
broken power--law model a better representation of these data.

In both cases, however, the spectral properties of the continuum 
for this object confirm its ``HBL--like'' character showed in the 
\sax observations, displaying steep spectra in all three datasets. 
This result is confirmed also accounting for the possible ROSAT miscalibration
(i.e., subtracting a maximum $\Delta\alpha \sim 0.2-0.3$). 
The broken power--law fits, which show a flattening of the spectrum towards 
higher energies, may indicate  that the inverse Compton contribution is becoming not
negligible, i.e., that the region where the synchrotron and Compton emissions 
cross was only slightly beyond the ROSAT band.

\subsection{RGB J1722$+$2436} 

This source is included in the RASS. \citet{yua98} give a $0.1-2.4$ keV
flux $\sim 4.2 \times 10^{-12}$ erg cm$^{-2}$ s$^{-1}$, 
i.e., a factor $\sim 4$ larger than the \sax
value, with a spectral index derived from the hardness ratio $\alpha_{\rm x} =
1.27^{+0.26}_{-0.28}$, assuming Galactic absorption. This is steeper than the
\sax value and again suggests a ``steeper when brighter'' behavior.

\subsection{Summary}\label{ROSAT_sum}

To summarize our ROSAT spectral results, the fitted energy indices are
relatively steep, $\alpha_{\rm x} \sim 1.4$, steeper than the \sax values by
$\Delta \alpha \approx 0.5$. Possible ROSAT miscalibrations are supposed to
account for a steepening of the order $\Delta \alpha \sim 0.2 - 0.3$, so we
regard this difference as significant. In two cases the source was also
brighter, suggesting a ``steeper when brighter'' behavior, indicative of a
shift of the synchrotron peak towards higher energies during high states.
Overall, this indicates a non-negligible synchrotron component in the X-ray
band, which actually dominates for RGB J1629+4008. 

\section{Spectral Energy Distributions} 

To address the relevance of our \sax data in terms of emission processes we
have assembled multifrequency data for all our sources. The main source of
information was NED, and the data are not simultaneous with our \sax
observations. We also looked for near--infrared observations in the Two Micron
All Sky Survey (2MASS) Second Incremental Release \citep{cut00}, finding
data only for WGA J0546.6$-$6415. 

\subsection{Optical Variability}

Given that our fits are mostly based on X-ray and optical data and that the
latter are particularly important to determine $\nu_{\rm peak}$, we need to
address the non-simultaneity of the optical and X-ray observations. The
optical fluxes of FSRQ are in fact known to show significant variability,
which could influence our results. We have then investigated the range of
optical variability of our sources, with the following results. 

\begin{itemize}

\item{WGA J0546.6$-$6415.} This is a newly identified source from the DXRBS
(Perlman et al. 1998). Nevertheless, we did find variability information using
SuperCOSMOS Sky Survey \citep{ham01} data, spanning the period 1975 -- 1986.
In particular, the UK Schmidt telescope (UKST) and European Southern
Observatory (ESO) red observations, taken at the end of 1985/beginning of 1986
and separated by 50 days, agree within 0.01 magnitude, and give $R \sim 15.2$.
This is 0.5 magnitudes fainter than the $F=14.7$ magnitude given in the USNO A2.0
catalogue, which refers to mid 1980. The Guide Star Catalogue-II
(GSC-II), available at {\tt http://www-gsss.stsci.edu/}, gives $J =
15.9\pm0.2$ and $F = 15.3\pm0.2$ for 1989, which indicates a brightening of
$\sim 0.6$ magnitudes compared to the UKST blue value of 16.5 for 1975.

\item{RGB J1629$+$4008.} This is a Faint Images of the Radio Sky at Twenty
centimeters (FIRST) \citep{bec95} quasar. \citet{hel01} have studied the
long-term optical variability of FIRST quasars. They found that in 1953
Palomar Observatory Sky Survey (POSS I) observations this object was $\sim
0.4$ magnitudes fainter in $B$ and $\sim 0.6$ magnitudes fainter in $R$ than
in 1996 CCD observations. The GSC-II gives $J = 18.0\pm0.4$ and $F =
17.5\pm0.4$ for 1993. This is perfectly consistent with the CCD $B$ and $R$
values given by \citet{hel01}.

\item{RGB J1722$+$2436.} This source was identified as a quasar by
\citet{bon77} due to its optical variability. These authors quote a range of
photographic magnitude for this source between 15.7 and 16.7, with long
``still-stands'' between brief periods of activity. The GSC-II gives $J =
16.7\pm0.4$ and $F = 15.2\pm0.4$ for 1996, in good agreement with the $O$ and
$E$ APM values given in Tab. 1.

\item{S5 2116$+$81.} The GSC-II gives $J = 14.4\pm0.4$ for 1993, consistent
with the APM $O$ value given in Tab. 1.

\end{itemize}

In summary, within the available optical data, our sources display maximum
variability amplitude $\sim 0.5-1$ magnitudes. We have then conservatively
assigned to the optical fluxes an error of 0.2 dex, which
corresponds to $\pm 0.5$ magnitudes (see Fig. \ref{fig7}).

\subsection{Synchrotron Inverse Compton Model Fits} 

\begin{figure*}
\vskip -0.1truecm
\hspace{1.3cm}\includegraphics[width=16.0cm]{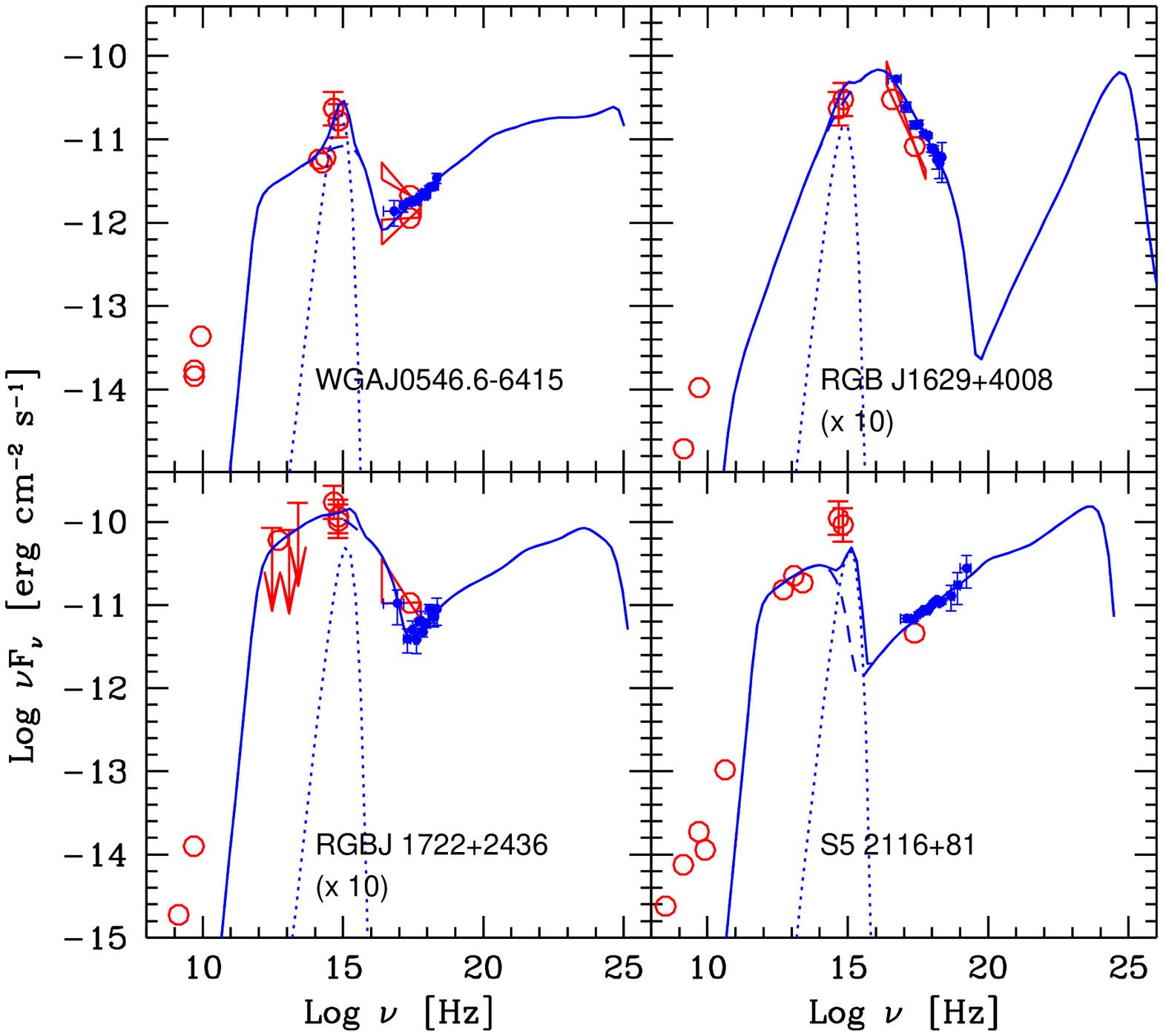}
\vskip -2truecm
\caption{Spectral energy distributions of our sources. Open points represent
data from the literature (NED) while the \sax data are indicated by filled
points. ROSAT data are shown by a bow--tie that represents the spectral index
range. The dashed lines correspond to the one--zone homogeneous synchrotron
and inverse Compton model calculated as explained in the text, with the
parameters listed in Table \ref{tab_sed}. Dotted lines represent the disk
emission, while solid lines indicate the sum of disk plus non-thermal
emission. The errors on the optical fluxes correspond to $\pm 0.5$ magnitudes
(see text for details).\label{fig7}}
\end{figure*}

The SEDs for our sources are shown in Figure
\ref{fig7}. The \sax data have been converted to $\nu f_{\rm \nu}$ units using the
XSPEC unfolded spectra after correcting for absorption. ROSAT data are
shown by a bow--tie that represents the spectral index range. 

To derive the intrinsic physical parameters which could account for the
observed data we have fitted the SED of our sources with a homogeneous,
one--zone synchrotron inverse Compton model as developed in \citet{ghi02}.
This model is very similar to the one described in detail in Spada et
al. (2001; it is the ``one--zone'' version of it), and is characterized by a
finite injection timescale, of the order of the light crossing time of the
emitting region (as occurs, for example, in the internal shocks scenario,
where the dissipation takes place during the collision of two shells of fluid
moving at different speeds).
In this model, the main emission
comes from a single zone and a single population of electrons, with the
particle energy distribution determined at the time $t_{\rm inj}$, i.e., at the
end of the injection, which is the time when the emitted luminosity is
maximized. Details of the model can be found in
\citet{ghi02}, who have applied it successfully to both low--power BL
Lacs and powerful FSRQ. Here we summarize its main characteristics.

The source is assumed cylindrical, of
radius $R$ and width $\Delta R^\prime = R/\Gamma$ (in the comoving frame, where
$\Gamma$ is the bulk Lorentz factor). The particle distribution $N(\gamma)$
is assumed to have the slope $n$ [$N(\gamma)\propto \gamma^{-n}$] above the
random Lorentz factor $\gamma_{\rm c}$, for which the radiative cooling time
equals the injection time. The latter is assumed to be equal to the light
crossing time (i.e., $t_{\rm inj}\sim \Delta R/c$). The electron distribution
is assumed to cut--off abruptly at $\gamma_{\rm max}>\gamma_c$. Below
$\gamma_c$ there can be two cases, depending on the values of $\gamma_c$ and
$\gamma_{\rm min}$:

i) If $\gamma_c >\gamma_{\rm min}$, 
we have $N(\gamma) \propto \gamma^{-(n-1)}$ between
$\gamma_{\rm min}$ and $\gamma_{\rm c}$ and 
$N(\gamma)\propto \gamma^{-1}$ below $\gamma_{\rm min}$. 

ii) Alternatively, if 
$\gamma_c<\gamma_{\rm min}$,
then $N(\gamma) \propto \gamma^{-2}$ between $\gamma_{\rm c}$ 
and $\gamma_{\rm min}$ and $N(\gamma) \propto \gamma^{-1}$ 
below $\gamma_{\rm c}$. 

According to these assumptions, the random Lorentz factor $\gamma_{\rm peak}$
of the electrons emitting most of the radiation (i.e., emitting at the 
peaks of the SEDs) is determined by the 
relative importance of radiative losses and can 
assume values in the range  $\gamma_{\rm min}$ to $\gamma_{\rm max}$.

Our sources have broad lines and therefore the contribution of the disk to the
SED could be relevant. 
Furthermore, photons produced in the broad line region could contribute
to the seed photon distribution for the inverse Compton scattering.
We accounted for this by assuming that a fraction $\sim$ 10\%
of the disk luminosity $L_{\rm disk}$ is reprocessed into line
emission by the broad line region (BLR), $L_{\rm BLR}$, assumed to be located
at $R_{\rm BLR}$. $L_{\rm BLR}$ was estimated following \citet{cel97}
from the fluxes of the strongest broad lines. These were
available for all our sources from our own spectrum (WGA J0546.6--6415;
\citet{per98}), from published spectra kindly made available to us by
Sally Laurent--Muehleisen (RGB J1629$+$4008 and RGB J1722$+$2436) and from
\citet{sti93} (S5 2116$+$81). $R_{\rm BLR}$ is assumed to
scale as $L_{\rm disk}^{0.7}$, following \citet{kas00}. Disk emission is
assumed to be a simple black--body peaking at $10^{15}$ Hz. Within these
assumptions, we can fix two important parameters for our modeling, and we can
reliably estimate the importance of the seed photons produced externally to
the jet for the Compton scattering process.

The source is assumed to emit an intrinsic luminosity $L^\prime$ and to be
observed with the viewing angle $\theta$. The model parameters are listed in
Table \ref{tab_sed}, which gives the name of the source in column (1),
$L^\prime$ in column (2), $L_{\rm disk}$ in column (3), $R_{\rm BLR}$ in
column (4), the magnetic field $B$ in column (5), the size of the region $R$
in column (6), the Lorentz factor $\Gamma$ in column (7), the viewing angle
$\theta$ in column (8), the slope of the particle distribution $n$ in column
(9), the minimum Lorentz factor of the injected electrons $\gamma_{\rm min}$
in column (10), and $\gamma_{\rm peak}$ and $\nu_{\rm peak}^{\rm syn}$
in columns (11) and (12) respectively. Column (13) gives the magnetic field
$B_{\rm eq}$ derived assuming equipartition of the energy content of the
magnetic field with the electron energy. Note that $L^\prime$, $B$, $R$,
$\Gamma$, $\theta$, $\gamma_{\rm min}$ and $n$ are varied during the
fitting procedure, $L_{\rm disk}$ and $L_{\rm BLR}$ are fixed by observations,
while $\gamma_{\rm peak}$, $\nu_{\rm peak}^{\rm syn}$, and $B_{\rm eq}$ are derived
quantities.

In the case of a pure synchrotron self--Compton model, all the above
parameters are constrained in sources for which: 1) we have an estimate of the
minimum timescale of variability; 2) both the synchrotron and the
self--Compton peak are well defined; 3) the spectral slopes below and above
the peaks are known; 4) the redshift is known. As discussed in \citet{tav98}, 
this suffices to fix the values of the magnetic
field, the intrinsic power of the source, the slopes of the emitting electron
distribution, the relativistic Doppler factor, and the dimension of the
source. When the radiation produced externally to the jet is important there
is one unconstrained unknown, 
but the superluminal motion of the radio knots
observed in blazars indicates values of the bulk Lorentz factor in the range
10--15 on average, and we therefore use these values for our fits (see, e.g.,
Ghisellini et al. 1998).

We do not have information about the high--energy (Compton) peak for the
sources in our sample so we lack the direct determination of $\gamma_{\rm
peak}$. But in the model we use here the finite time of injection of particles
plays a crucial role, and we have an additional constraint with respect to the
simplest synchrotron inverse Compton model. Namely, the peak of the
synchrotron emission is either due to the electrons injected with $\gamma_{\rm
min}$, or it is due to the electrons which were able to radiatively cool in
the timescale $t_{\rm inj}$ (i.e., electrons with Lorentz factor $\gamma_c$).
In the latter case (which is verified for all the sources studied in this
paper but RGB J1629$+$4008), the value of $\gamma_c$ depends on the total
energy density (magnetic plus radiative) as seen in the comoving frame. Since
the ``external'' radiation energy density is known once we specify the bulk
Lorentz factor (it is assumed to be produced by the BLR), $\gamma_c$ depends
on the magnetic field. The break in the particle distribution at $\gamma_c$
corresponds to a break of the emitted synchrotron spectrum (and can often
correspond to the peak of the synchrotron spectrum) and in this way we can
reliably estimate the value of the magnetic field.

\begin{table*}
\vskip -2truecm
{\footnotesize
\begin{center}
\caption{Model Parameters.\label{tab_sed}}
\begin{tabular}{lllllllllllll}
\tableline\tableline
Name & $L^\prime$ & $L_{disk}$ & $R_{BLR}$ & $B$ & $R$ & $\Gamma$ & $\theta$ & $n$ &
$\gamma_{min}$ & $\gamma_{peak}$ & $\nu^{syn}_{peak}$ & $B_{\rm eq}$ \\
 & erg s$^{-1}$ & erg s$^{-1}$ & cm & G & cm
&    &    &   &     & &Hz &G  \\
\tableline
WGA J0546.6$-$6415 &2.5e42  &1.4e46  &2.7e18  &1.0  &8.0e15 &12  &3.5  &3.6  &3.0e1 &1.0e4  &5.8e15 
&3.2\\
RGB J1629$+$4008   &3.0e41  &7.2e44  &3.5e17  &2.5  &1.0e16 &13  &3.0  &4.2  &1.0e4 &1.0e4  &1.7e16 
&0.9\\
RGB J1722$+$2436   &1.3e42  &9.3e44  &4.0e17  &4.0  &7.0e15 &13  &5.0  &3.5  &1.0e2 &1.5e3  &3.8e14 
&2.7\\
S5 2116$+$81       &2.0e42  &1.9e45  &7.0e17  &1.0  &1.0e16 &11  &5.0  &3.5  &2.0e1 &3.0e3  &3.9e14 
&2.3\\
\tableline
\end{tabular}
\end{center} }
\end{table*}

We are aware of the fact that for three of our objects the shape, sampling,
and lack of simultaneity of the SED do not allow us to firmly constrain the
peak of the synchrotron emission (the exception being RGB J1629+4008), also
because of the non-negligible contribution of the thermal disk component to
the optical flux (especially for WGA J0546.6$-$6415 and S5 2116$+$81; see
Fig. \ref{fig7}). However, some reasonable arguments can be made. For WGA
J0546.6$-$6415 and RGB J1722$+$2436 the presence of a steeper (and variable)
component at soft X-ray energies (\S\S~\ref{SAX_sum} and \ref{ROSAT_sum}) 
may be attributed to the tail of the
synchrotron emission (see also Sect. \ref{bump}), thus suggesting a
synchrotron peak which cannot be at too low energies ($\ga 10^{14}$ Hz). This
is corroborated also by the IR and optical data for RGB J1722$+$2436, whose
optical flux is not dominated by the disk emission. In the case of WGA
J0546.6$-$6415, which has a stronger disk component, the synchrotron peak
could reach $\sim 10^{13}$ Hz only if one neglects the evidence for an upturn at
low X-ray energies. S5 2116$+$81 is the less constrained source, but it is
also the case for which we have no indication of a steep, soft X-ray
component. The radio to IR data suggest a peak above $10^{13}$ Hz and no set
of parameters was found to be compatible with the IR and \sax data and with a
$\nu_{\rm peak}^{\rm syn}$ less than $\sim {\rm a~few} \times 10^{13}$ Hz. 
For all three
objects, however, the synchrotron peak frequency has to be $<10^{16}$ Hz,
since otherwise the X-ray band would have a strong and obvious steep
synchrotron emission (as for RGB J1629$+$4008), contrary to what observed. In
summary, we have modelled our objects mainly on the basis of the \sax (and partly
ROSAT) data,
trying to fit the other data as far as they could be compatible with the \sax
ones. The derived parameters, then, apart from RGB J1629+4008, should be
regarded as tentative.

The model fits are shown in Fig. \ref{fig7} as solid lines. The applied model
is aimed at reproducing the spectrum originating in a limited part of the jet,
thought to be responsible for most of the emission. This region is necessarily
compact, since it must account for the fast variability shown by all blazars,
especially at high frequencies. The radio emission from this compact region is
strongly self--absorbed, and the model cannot account for the observed radio
flux. This explains why the radio data are systematically above the model fits
in the figure. Our model fails also to reproduce the optical flux of S5
2116+81. This is the source with the lowest redshift ($z=0.084$) and we
therefore expect a non--negligible contamination from the host galaxy, which
is clearly detected in the Digitized Sky Survey (DSS) plates.

Note that the model fits predict that the high energy inverse Compton emission
has a luminosity comparable to the synchrotron one in all sources but one, S5
2116+81. The shape of the inverse Compton component is sometimes complex,
because of the contributions of the synchrotron self Compton component,
especially in the X--ray range, and the external component, especially at
higher energies.

As shown in Table \ref{tab_sed}, the intrinsic luminosities, the source
dimensions, the bulk Lorentz factors and the viewing angles are quite similar
for all sources. The fact that RGB J1629+4008 has a steep X--ray
(synchrotron) spectrum is explained by our model by the very large value of
$\gamma_{\rm min}$, the minimum Lorentz factor of the injected electrons which
in this case corresponds to $\gamma_{\rm peak}$. The inferred values of the magnetic
field are not very far from the equipartition values, with an average ratio 
$\langle B/B_{\rm eq} \rangle \sim 1.3$.

\section{Discussion}

Our source selection requires the objects to be in the HBL region and
therefore to have relatively low $\alpha_{\rm ro}$ values. Given also their
relatively low radio fluxes (and powers; see \S~\ref{interpr}), before
discussing our results we need to assess if our sources are of the same kind
as the more powerful and more studied blazars. We do this in the following
subsection.

\subsection{The Nature of Our Sources}

\subsubsection{Are our sources radio--loud?} 

Two different definitions of radio--loud quasars have been used in the
literature (see, e.g., Stocke et al. 1992). The first is based on the
rest--frame ratio $R$ of radio (typically 5 GHz) to optical (typically 4,400
\AA) flux density, $\log R = \log(f_{\rm r}/{\rm mJy}) +0.4B-6.66$
(independent of redshift if $\alpha_{\rm r} = \alpha_{\rm o} = 0.5$ is
assumed). Radio--loud sources are taken to have $R > 10$ which translates, in
our notation, to $\alpha_{\rm ro} \ga 0.2$. The other definition is based on
radio power and appears to be redshift-dependent, with a dividing line ranging
from $\approx 10^{24}$ W Hz$^{-1}$ for the Palomar Green (PG) sample
(optically bright and at relatively low-redshift) up to $10^{26}$ W Hz$^{-1}$
for the Large Bright Quasar Survey (LBQS) sample (optically fainter and at
higher redshift). The two definitions overlap somewhat \citep{pad93}. As
shown in Fig. \ref{fig1} and Tab. 1, all our sources are firmly in the radio-loud
regime as far as the first definition is concerned. All of them have
$\alpha_{\rm ro} \ge 0.26$, with $\langle \alpha_{\rm ro} \rangle \sim 0.33$
(or $\langle R \rangle \sim 50$; typical radio-quiet sources [i.e., optically
selected quasars] have $\alpha_{\rm ro} \sim 0.1$ or $R \sim 0.5$). The 5 GHz
radio powers of our sources are in the range $10^{24.6} - 10^{26}$ W
Hz$^{-1}$, with $\langle L_{\rm r} \rangle \sim 10^{25.1}$ W Hz$^{-1}$. Given
their relatively low redshift ($\langle z \rangle \sim 0.2$), these values
also classify our sources as radio-loud.

\subsubsection{Are our sources blazars?}\label{blazars?}

As described in the Introduction, blazars are characterized
by a variety of properties, which have also led to a proliferation of names
(HPQ, OVV, CDQ, FSRQ). The flat--spectrum radio quasar definition is the
easiest one to apply, and it appears that the majority of powerful FSRQ indeed
show rapid variability, high polarization, and radio structures dominated by
compact radio cores, and vice versa (Urry \& Padovani 1995 and references
therein).

As our sources are less powerful than the FSRQ commonly studied (see
\S~\ref{powers}), it is important to check, as far as possible, that we
are dealing with the same type of sources. We do this in the following. 

Our objects have, by selection, a flat radio spectrum. In fact, $\langle
\alpha_{\rm r} \rangle = -0.37$, that is their typical radio spectrum is not
only flat but inverted. Three of our sources are included in the NVSS and
therefore low-resolution (45$^{\prime\prime}$) radio maps at 1.4 GHz are
available. RGB J1629$+$4008 and RGB J1722$+$2436 are compact and
unresolved. RGB J1629$+$4008 is also included in the FIRST \citep{bec95} 
radio survey, which has a
higher resolution (5$^{\prime\prime}$) than the NVSS. This source is still
compact and unresolved in the FIRST map, with a 1.4 GHz radio flux of
$11.9\pm0.2$ mJy, as compared to $9.0\pm0.5$ mJy in the NVSS. The fact that
the higher-resolution FIRST flux is similar to (actually larger than) the NVSS
flux suggests that the source is core-dominated (modulo variability
effects). S5 2116$+$81 has a complex NVSS map, with four radio sources within
3\amin from the source position. However, dedicated radio observations
\citep{tay96} show that the VLA and VLBA (Very Long Baseline Array) 5 GHz core
fluxes of the stronger source are basically the same, again indicating a very
compact core. Finally, our own 4.8 and 8.6 GHz Australia Telescope Compact
Array (ATCA) maps of WGA J0546.6$-$6415 (2$^{\prime\prime}$ and
1$^{\prime\prime}$ resolution, respectively) indicate a compact, unresolved
source, with a core-dominance parameter at 4.8 GHz $\la 35$ (see Landt et
al. 2001 for details of the ATCA observations). The fact that the
higher-resolution ATCA flux at 5 GHz is similar to (actually larger than) the
lower-resolution Parkes-MIT-NRAO (PMN) \citep{wri94} flux (339 mJy vs. 287 mJy) again
suggests that the source is core-dominated (modulo variability effects).

Optical polarization data are available only for two of our sources, RGB
J1722$+$2436 and S5 2116$+$81. The former has $P_{\rm opt} = 0.70\pm0.48$\% 
\citep{sto84} while the latter has $P_{\rm opt} = 0.94 \pm0.21$\%
\citep{mar96}. We note that although these values are below the ``canonical''
division between high-polarized (HPQ) and low-polarized (LPQ) quasars (and are
indeed not inconsistent with possible scattering by the interstellar medium of
our own galaxy), many flat--spectrum LPQ are also thought to be seen with
their jets at small angles w.r.t. the line of sight (e.g., Ghisellini et
al. 1993), and therefore classify as ``blazars'' according to our
definition. Furthermore, if indeed these objects are the equivalent to HBL, as
discussed in \S~7.2, we would expect their duty-cycle for high polarization to
be lower than their low-frequency peaked relatives, as was found for HBL and
LBL by \citet{jan94}. Finally, the relative prominent
thermal emission, especially in S5 2116$+$81 (see Fig. \ref{fig7}), would also
explain a low optical polarization.

As regards variability, at least three of our sources appear to vary in the
X--ray band, based on our \sax and ROSAT comparison, with RGB J1629+4008
decreasing in flux by a factor of 4 in 7 hours in our MECS and LECS observations. 
 
\subsection{Have we found high-frequency peaked FSRQ?}

The purpose of this paper was to study the X--ray properties and, more
generally, the SED of a class of flat--spectrum radio quasars with relatively
large X--ray--to--radio flux ratios and effective spectral indices typical of
HBL. We now address the question if these sources are really the strong,
broad-lined counterparts of HBL. 

\subsubsection{Spectral Energy Distributions}

The SED of RGB J1629+4008 is HBL--like, with the steep X--ray spectrum clearly
attributable to synchrotron emission. For WGA J0546.6$-$6415 and RGB
J1722+2436 the situation is less clear. Our fits and the SEDs are not
inconsistent with the fact that the \sax band might be close to the ``valley''
between synchrotron and inverse Compton emission, which would classify them as
akin to ``intermediate'' BL Lacs. Finally, S5 2116+81 appears to be LBL--like.

\subsubsection{X--ray spectral index and the synchrotron peak frequency}

A good indicator of the dominant emission process in the X--ray band is the
X--ray spectral index -- synchrotron peak frequency plot \citep{pad96,lam96}.
\citet{pad96} found a strong anti--correlation between the ROSAT $\alpha_{\rm
x}$ and $\nu_{\rm peak}$ for HBL (i.e., the higher the peak frequency, the
flatter the spectrum), while basically no correlation was found for LBL. This
was interpreted as due to the tail of the synchrotron component becoming
increasingly dominant in the ROSAT band as $\nu_{\rm peak}$ moves closer to
the X--ray band (see Fig. 7 of Padovani \& Giommi 1996).

The \sax version of this dependence was studied by Padovani et al. (2001; see
their Fig. 7). They confirmed the ROSAT findings, namely a strong
anti-correlation between $\alpha_{\rm x}$ and $\nu_{\rm peak}$ for HBL and no
correlation for LBL, with an initial increase in $\alpha_{\rm x}$ going from
LBL to HBL (see Padovani et al. 2001 for details).

\centerline{\includegraphics[width=9.0cm]{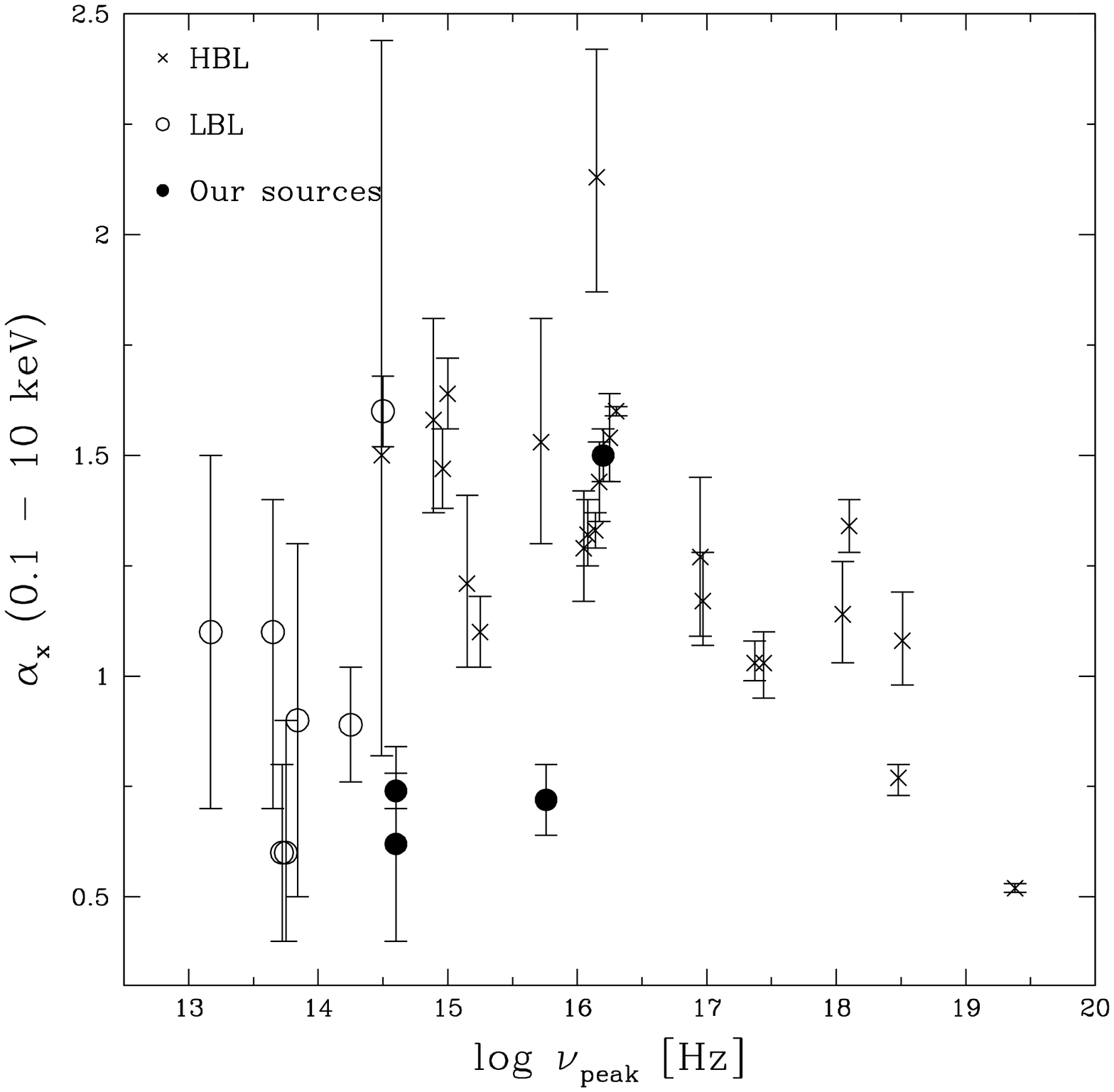}}
\figcaption{The {\it BeppoSAX} spectral index ($0.1-10$ keV) versus the logarithm
of the synchrotron peak frequency for our sources (filled circles), HBL
(crosses), and LBL (open circles). Data for LBL come from \citet{padet01},
while those for HBL come from Wolter et al. (1998; updated in Beckmann et
al. 2002), \citet{bec02}, and \citet{padet01}. Three of our sources, namely
those with values of $\alpha_{\rm x}< 1$, have quite uncertain $\nu_{\rm peak}$
values but likely in the range $10^{14} \la \nu_{\rm peak} < 10^{16}$ Hz 
(see text for details).\label{fig8}}
\centerline{}
\vskip 0.2in


Fig. \ref{fig8} shows an updated version of the $\alpha_{\rm x} - \nu_{\rm
peak}$ plot for BL Lacs from \citet{padet01}, including the HBL studied by
\citet{bec02}. We have also included in this plot our sources, represented by
filled points. We notice that RGB J1629+4008 falls right in the region
occupied by HBL, again confirming our interpretation that its \sax band is
dominated by synchrotron emission. This is further supported by the detection
of rapid variability, which has been seen in the synchrotron emission of
several BL Lacs observed with \sax but never in the inverse Compton component,
which varies on much longer time scales (e.g, \citet{rav02} and references 
therein). The other three sources have
values which, although quite uncertain, are probably in the
range $10^{14} \la \nu_{\rm peak} < 10^{16}$ Hz, which overlap with the values 
typical of
``intermediate'' BL Lacs, but display relatively flat $\alpha_{\rm x}$. We notice that
the scatter in the diagram is larger in the $\nu_{\rm peak}$ range $\approx
10^{14} - 10^{15.5}$ Hz, which is also where the HBL/LBL division becomes
blurrier. We interpret this as due to the fact that in this range of $\nu_{\rm
peak}$ objects can be both synchrotron and inverse Compton dominated,
depending on variability state.  For comparison, note that 3C 279, a classical
FSRQ, has a \sax spectral index $\alpha_{\rm x} = 0.66\pm0.04$ and $\nu_{\rm
peak} \sim 10^{13}$ Hz \citep{bal02}, i.e., well into the LBL region.

\subsubsection{The ``Blue Bump'' Component}\label{bump} 

Fig. \ref{fig7} shows that the thermal component is non--negligible in most of
our sources, with a ratio of black--body to synchrotron emission at $\sim
5,000$ \AA~ranging from $\sim 2$ for WGA J0546.6$-$6415 and S5 2116$+$81 to
$\sim 0.2$ for RGB J1722$+$2436. While these ratios are model--dependent,
especially for the former sources for which the optical band is close to the
synchrotron cut--off, it is instructive to check the effect of the thermal
component on the position of our objects in the $\alpha_{\rm ro} - \alpha_{\rm
ox}$ plane. By subtracting the thermal component and therefore using only the
synchrotron flux in the optical band to derive effective spectral indices, we
can see where the objects would fall (we assume that the blue bump has no
effect at 1 keV; but see below). As $\alpha_{\rm rx}$ stays constant while
$\alpha_{\rm ro}$ and $\alpha_{\rm ox}$ get steeper and flatter
respectively, the sources move along diagonal lines up and to the left in
the diagram. More specifically, the ``synchrotron only'' values are as
follows: WGA J0546.6$-$6415, $\alpha_{\rm ro} \sim 0.5$, $\alpha_{\rm ox} \sim
1$, RGB J1629$+$4008, $\alpha_{\rm ro} \sim 0.4$, $\alpha_{\rm ox} \sim 0.85$,
S5 2116$+$81, $\alpha_{\rm ro} \sim 0.4$, $\alpha_{\rm ox} \sim 1.2$
(practically no change is required for RGB J1722$+$2436). It then follows that
even by subtracting the effect of the thermal component our sources would
still be well within the HBL region in the $\alpha_{\rm ro} - \alpha_{\rm ox}$
plane (Fig. \ref{fig1}). Actually, the effect of a strong blue bump, by
moving sources to the right and down in the diagram, is that of ``pushing''
objects off the HBL region and into the radio-quiet zone. 
Even with some freedom in the modelling of the synchrotron component,
the location of our sources in the HBL region results quite robust, and 
cannot be ascribed to the accretion disk. 
That relatively high values of $\nu_{\rm peak}$ are not due to the accretion disk
is also indicated by the low, ``HBL--like'' values of $\alpha_{\rm rx}$,
a parameter well correlated with $\nu_{\rm peak}$ (e.g., Padovani \& Giommi
1996) and not affected by the optical contamination.   
Note that the
$\nu_{\rm peak}$ values given in Tab. \ref{tab_sed} and shown in
Fig. \ref{fig8} are those relative only to the synchrotron component and therefore
do not include the thermal component. 

In Seyfert galaxies and radio--quiet quasars the X--ray emission is thought
to be produced by a hot thermal corona sandwiching a relatively cold accretion
disk. One could then ask if the same component can contribute also in our
objects, two of which indeed have a relatively large disk component in the 
optical--UV band. 
While it is certainly possible that some fraction of the
X--ray emission (especially at low energies) is produced by a hot corona, 
we believe that the non--thermal 
components in these objects are dominant, due to the X-ray variability
indicated by the comparison between \sax and ROSAT data, particularly
the rapid LECS/MECS variability seen in RGB J1629+4008, and to the
lack of any indication of iron line features and hardening of the spectrum due
to the so--called Compton hump (which are instead common in Seyfert galaxies).
A possible way to decrease the importance of the iron line emission and Compton hump
in the spectrum could be to assume a dynamic corona, but in such case the corona 
emission would be characterized by harder X-ray spectra ($\alpha_{\rm x} < 1$; 
see Malzac et al. 2001), and thus could not account for soft excesses or steep 
spectra.
  
\subsubsection{Source Powers}\label{powers}

As discussed above, the 5 GHz radio powers of our sources are in the range
$10^{24.6} - 10^{26}$ W Hz$^{-1}$, with $\langle L_{\rm r} \rangle \sim
10^{25.1}$ W Hz$^{-1}$. Their 1 keV X--ray powers are in the range $10^{19.4}
- 10^{20.6}$ W Hz$^{-1}$, with $\langle L_{\rm x} \rangle \sim 10^{20.1}$ W
Hz$^{-1}$. These values, although not very different from the mean values of
the EMSS BL Lacs ($\langle L_{\rm r} \rangle \sim 10^{24.8}$ W Hz$^{-1}$ and
$\langle L_{\rm x} \rangle \sim 10^{20.3}$ W Hz$^{-1}$), are still in the FSRQ
range, albeit on the low side (see, for example, Fig. 5 of Perlman et al. 1998
and Padovani et al., in preparation). Their radio powers put our sources near the
low--luminosity end of the DXRBS FSRQ radio luminosity function \citep{pad01}.

\subsection{Astrophysical Interpretation}\label{interpr} 

We have found three examples of strong, broad--lined blazars with an SED
typical of HBL (one source) and intermediate BL Lacs (two sources). While the
first case is quite strong, the other two are more tentative. These are the
first such sources, as most broad--lined blazars known so far have SED typical
of LBL, i.e., with synchrotron peaks at IR/optical energies.

Our FSRQ, however, have relatively low powers, more typical of BL Lacs than
quasars. This ties in with the suggestion by \citet{ghi98} and \citet{ghi02}
that blazars form a sequence, controlled mainly by
their bolometric luminosity which in turn controls the amount of cooling
suffered by the emitting electrons. In powerful sources, where cooling is
severe (high radiative ($U_{\rm r}$) plus magnetic field ($U_{\rm B}$) energy 
densities), electrons of all energies cool rapidly (i.e., in a time shorter than
the injection time), making the particle distribution to have a break at
$\gamma=\gamma_{\rm min}$, which in this case becomes $\gamma_{\rm peak}$.
These sources are typically LBL--like, and are those located in the lower
part of the $\gamma_{\rm peak} - U_{\rm r}+U_{\rm B}$ plot (see Fig. \ref{fig9}).
In less powerful
sources, instead, only the highest energy electrons can cool in the injection
time, and consequently the particle distribution will have a break at $\gamma=
\gamma_{\rm c}$, for which $t_{\rm cool}=t_{\rm inj}$. In this case
$\gamma_{\rm peak}=\gamma_{\rm c}$ (and $\propto (U_{\rm r}+U_{\rm B})^{-1}$;
see details in \citet{ghi02}).
Fig. \ref{fig9} shows that the FSRQ studied in this paper, which have relatively
high $\gamma_{\rm peak}$ (and therefore $\nu_{\rm peak}$) values, appear to
fit within this scenario, as they follow the suggested sequence. (Note that,
given the scatter in the plot, this conclusion stands even in the case of a
decrease of $\sim$ one order of magnitude in the $\gamma_{\rm
peak}$ values for the three sources with relatively uncertain
estimates of this parameter.) Moreover, in agreement with
their relatively low powers, they are positioned at the low-end of the FSRQ
region and actually within the BL Lac region. In other words, we still have
not found a case of high--$\nu_{\rm peak}$ (and therefore high--$\gamma_{\rm
peak}$) {\it and} high--power blazar, which would invalidate this scenario.

\centerline{\includegraphics[width=10.5cm]{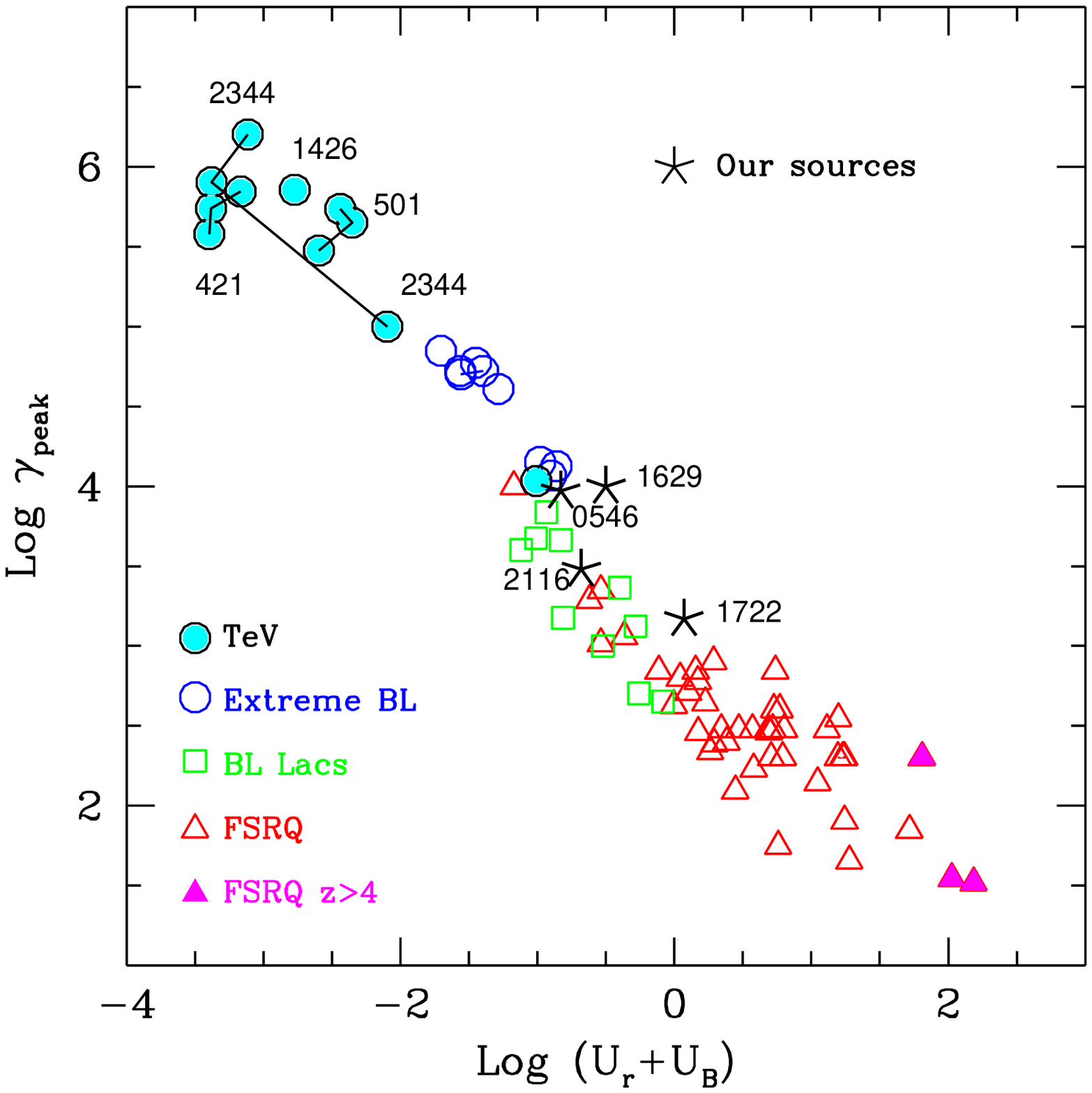}}
\vskip -0.8truecm
\figcaption{$\gamma_{\rm peak}$, the Lorentz factor of the electrons emitting
most of the radiation, vs. the total energy density in the emitting region,
i.e., the sum of the radiative ($U_{\rm r}$) and magnetic field ($U_{\rm B}$)
energy densities, as measured in the comoving frame. Open circles represent
``extreme'' HBL, filled circles represent TeV sources, open squares represent
BL Lacs, open triangles represent FSRQ, filled triangles are FSRQ at $z>4$,
while stars indicate our sources (HFSRQ). All data apart from the latter are
from \citet{ghi02}, updated in part from \citet{ghi98}. 
\label{fig9}}
\centerline{}
\vskip 0.2in


Another explanation for the relatively low power of our sources is related to
a selection effect. Fig. \ref{fig7} shows that our only ``true'' HBL--like
FSRQ, RGB J1629$+$4008, has a synchrotron peak at $\sim 2 \times 10^{16}$ Hz,
well above the peak of the thermal emission. Its non-thermal power is also low
enough (as implied by its radio power $L_{\rm 5 GHz} \sim 6 \times 10^{24}$ W
Hz$^{-1}$) that the thermal component, and its associated broad lines, is
non--negligible in the optical/UV band. Given that the equivalent width of its
strongest emission line is $\sim 80$ \AA, the same object with a non-thermal
power a factor of $\sim 20$ higher, corresponding to $L_{\rm 5 GHz} \sim
10^{26}$ W Hz$^{-1}$, still rather modest for an FSRQ, would be classified as
a BL Lac object, since its equivalent width would go down to $\sim 4$ \AA. An
even stronger non-thermal component would completely wash out the emission
lines, making the redshift determination of this object impossible, as there
are no galaxy absorption features in its spectrum. It then follows that a
broad-lined source with the same SED as RGB J1629$+$4008 but a much larger
power would not even be classified as an FSRQ but as an HBL without
redshift. This could also explain why high--power, high--$\nu_{\rm peak}$ BL
Lacs have not been yet found. A key ingredient of this argument is the
frequency of the synchrotron peak. Assuming a typical $\nu_{\rm peak}$ value
for a ``classical'' FSRQ of $\sim 10^{13.5}$ Hz, which implies a much weaker
optical synchrotron component, an object with a radio power $\sim 100$ times
larger than RGB J1629$+$4008 will have the same optical (non-thermal) power
and therefore will still be recognized as a quasar (assuming similar thermal
powers). It follows that high--power, low--$\nu_{\rm peak}$ FSRQ can be
identified as quasars up to very high luminosities. Conversely,
high--$\nu_{\rm peak}$ blazars would be classified as HBL without redshift
already at moderate radio luminosities. One problem with this explanation for
the lack of high--$\nu_{\rm peak}$ --- high--power blazars is that there
should be intermediate cases, i.e., HBL with broad (but weak) lines. This does
not seem apparent, for example, from the optical spectra of the
EMSS BL Lacs (e.g., Rector et al. 2000).

As the selection of our sources was done $3-4$ years ago, one could ask if our
sources are still really extreme in terms of their X--ray--to--radio flux
ratios. Out of the 199 DXRBS FSRQ known to date, WGA J0546.6$-$6415 is still the
second most extreme. The FSRQ RGB sample, given its higher X--ray flux limit
and slightly lower radio flux limit, is even better suited to find sources
with high $f_{\rm x}/f_{\rm r}$ ratios (lower $\alpha_{\rm rx}$; see Padovani
et al., in preparation). Sorting the known RGB FSRQ in order of ascending
$\alpha_{\rm rx}$ value, RGB J1629$+$4008 is at the top of the list while RGB
J1722$+$2436 is number four (RGB J1413+4339, which is number two, is a
borderline quasar; see Laurent-Muehleisen et al. 1998). Finally, S5 2116$+$81
is number 15 out of the $\sim 400$ FSRQ in the multifrequency catalog of
\citet{pad97} with X--ray data sorted in order of ascending
$\alpha_{\rm rx}$ value. It was selected because it is the X--ray brightest.

We find it significant that, having selected some of the few known FSRQ with
effective spectral indices typical of HBL and therefore relatively large
X--ray--to--radio flux ratios, we found only one case where the X--ray band is
clearly dominated by synchrotron emission, with a $\nu_{\rm peak} \sim 2
\times 10^{16}$ Hz, while the other three sources have $\langle \nu_{\rm peak}
\rangle \la 10^{15}$ Hz. As shown, for example, in Fig. \ref{fig8},
HBL can reach much higher values, up to $10^{18} - 10^{19}$ Hz, albeit in
cases of extreme variability (e.g., MKN 501; Pian et al. 1998). A more
meaningful comparison is with the HBL in the multifrequency catalog of
\citet{pad97} having radio flux $> 20$ mJy, the minimum value for our
sources. In that case the $\nu_{\rm peak}$ distribution has a {\it typical}
value $\sim 4 \times 10^{16}$ Hz and reaches $\sim 2 \times 10^{17}$ Hz. A
discussion of the $\nu_{\rm peak}$ distribution for FSRQ and BL Lacs belonging
to the {\it same} sample will be presented by Padovani et al. (in preparation).

\citet{sam00} reported on {\it ASCA} observations of four FSRQ
characterized by steep ROSAT spectra ($\alpha_{\rm x} \sim 1.3$). The sources
were all found to have flat hard X--ray spectra, with $\alpha_{\rm x} \sim
0.8$. Sambruna et al. discuss their results in terms of relatively high
synchrotron peaks and thermal emission extending into the X--ray band. We
stress, however, that their sources sample a region of parameter space widely
different from ours. Their effective spectral indices, in fact, place their
FSRQ firmly in the LBL region, unlike ours, so that their four objects should
not have been expected to show high $\nu_{\rm peak}$ values and steep {\it
ASCA} spectra. \citet{geo00} used the fact that these four FSRQ happen
to have relatively low core-dominance parameters ($\approx 1$) compared to two
``intermediate'' BL Lacs to suggest that these sources could be seen at larger
angles than more typical FSRQ. We stress that, as discussed in
\S~\ref{blazars?}, our sources, and especially RGB J1629+4008, appear to be
quite compact so that his interpretation might not apply to our objects.

\section{Summary and Conclusions} 

We have presented new \sax observations of four flat--spectrum radio quasars
selected to have effective spectral indices ($\alpha_{\rm ro}$, $\alpha_{\rm
ox}$) typical of high-energy peaked BL Lacs. The main purpose of the paper was
to see if these sources are indeed the broad-lined counterparts of BL Lacs
with the synchrotron peak at UV/X--ray energies (HBL). Our objects are quite
extreme in terms of their X--ray--to--radio flux ratios (a factor $\sim 70$
higher than ``classical'' FSRQ), qualify as radio-loud sources both in terms
of their radio-to-optical flux ratio and power, and are variable and
radio-compact. Our main results can be summarized as follows:

\begin{enumerate} 

\item We have discovered the first FSRQ (RGB J1629+4008) whose X--ray emission
is dominated by synchrotron radiation, as clearly shown by our \sax
observations ($\alpha_{\rm x} \sim 1.5$ in the $0.1-10$ keV band), in which
the source was also rapidly variable. This object is therefore the first confirmed
member of a newly established class of HBL--like FSRQ. This result is fully
consistent with earlier ROSAT observations which detected a steep soft X--ray
spectrum ($\alpha_{\rm x} \sim 2$ in the $0.1-2.4$ keV band). The combination
of the X--ray data with archival radio and optical data gives a spectral energy
distribution typical of HBL sources, with a synchrotron $\nu_{\rm peak} \sim 2
\times 10^{16}$ Hz. The derived values follow very well the $\alpha_{\rm x} -
\nu_{\rm peak}$ correlation seen for HBL sources (Fig. \ref{fig8}).

\item The other three sources show a relatively flat ($\alpha_{\rm x} \sim
0.75$) \sax spectrum. However, we have found for two of them (WGA J0546$-$6415
and RGB J1722+2436) some indication of steepening at low-energies. This is
based on \sax and ROSAT data and on their (sparsely sampled) SEDs. We interpret this 
as the tail
of synchrotron emission, which is not strongly constrained but would peak around 
$\approx 10^{15}$ Hz, as typical of ``intermediate'' BL Lacs 
for which the synchrotron and inverse Compton components meet within
the \sax band. The last source (S5 2116+81) turned out to be a more typical
FSRQ, with the \sax band fully dominated by inverse Compton emission
and a synchrotron $\nu_{\rm peak}<10^{15}$ Hz. 

\item By fitting a synchrotron inverse Compton model which includes the
contribution of an accretion disk, whose power we estimate from the broad-line
region luminosity, to the spectral energy distributions, we have derived
physical parameters (e.g., intrinsic power, magnetic field, etc.) for our
sources. The thermal component was found to be non-negligible in three out of
four objects, with ratios of thermal to synchrotron emission in the range
$0.6 - 2$ at 5,000 \AA. 

\item Although the original selection was based mostly on the effective 
broad-band (radio through X--ray) colors, which are in first approximation 
independent of
luminosity, all four candidates are at relatively low powers, more typical of
BL Lacs than of FSRQ and in any case close to the low-luminosity end of the
FSRQ radio luminosity function \citep{pad01}. We interpret this as due to
two, possibly concurrent, causes: an inverse dependence of $\gamma_{\rm peak}$
(and therefore $\nu_{\rm peak} \propto \gamma_{\rm peak}^2 \delta B$), the
Lorentz factor of the electrons emitting most of the radiation, on bolometric
power, due to the more sever cooling at work in more powerful sources; and a
selection effect, namely the fact that a high-power, HBL--like quasar will have
such a strong optical/UV non-thermal component that its emission lines will be
swamped and the object will be classified as a featureless BL Lac. 
\end{enumerate}

A better understanding of these rare sources and the confirmation of the
non-existence of high--$\nu_{\rm peak}$ --- high--power blazars will require
the selection and study of a larger sample of objects. 

\acknowledgments

We thank Sally Laurent-Muehleisen for providing us with the optical spectra of
the two RGB sources. LC acknowledges the STScI Visitor Program. EP
acknowledges support from NASA grants NAG5-9995 and NAG5-10109 (ADP) and
NAG5-9997 (LTSA). This research has made use of the USNOFS Image and Catalogue
Archive operated by the United States Naval Observatory, Flagstaff Station
({\tt http://www.nofs.navy.mil/data/fchpix/}), of the NASA/IPAC Extragalactic
Database (NED), which is operated by the Jet Propulsion Laboratory, California
Institute of Technology, under contract with the National Aeronautics and
Space Administration, of the STScI Digitized Sky Survey, and of data products 
from the Two Micron All Sky Survey,
which is a joint project of the University of Massachusetts and the Infrared
Processing and Analysis Center/California Institute of Technology, funded by
the National Aeronautics and Space Administration and the National Science
Foundation. The Guide Star Catalogue-II is a joint project of the Space
Telescope Science Institute and the Osservatorio Astronomico di Torino. Space
Telescope Science Institute is operated by the Association of Universities for
Research in Astronomy, for the National Aeronautics and Space Administration
under contract NAS5-26555. The participation of the Osservatorio Astronomico
di Torino is supported by the Italian Council for Research in Astronomy.
Additional support is provided by European Southern Observatory, Space
Telescope European Coordinating Facility, the International GEMINI project and
the European Space Agency Astrophysics Division.

\end{document}